\documentclass{aa}  
\usepackage{graphicx}
\usepackage{txfonts}

\usepackage{amsmath}	% Advanced maths commands
\usepackage{amssymb}	% Extra maths symbols
\usepackage[usenames]{xcolor}
\usepackage[normalem]{ulem}

\usepackage{pifont}
\usepackage{xcolor}

\usepackage{natbib}
\bibpunct{(}{)}{;}{a}{}{,} % to follow the A&A style
\newcommand{\cmark}{\textcolor{green!80!black}{\ding{51}}}
\newcommand{\xmark}{\textcolor{red}{\ding{55}}}

\usepackage{hyperref}
\begin{document}

  \title{Orbital dynamics in the GG~Tau~A system: investigating its enigmatic disc}
   \titlerunning{A misaligned disc in GG Tau} 

   \author{Claudia Toci
          \inst{1,2}\fnmsep\thanks{E-mail: claudia.toci@eso.org},
           Simone Ceppi\inst{3},
           Nicol\'as Cuello\inst{4},
           Gaspard Duch\^ene$^{4,5}$,
           Enrico Ragusa$^{6,7}$,
           Giuseppe Lodato\inst{3},
           Francesca Farina\inst{3},
           Fran\c cois M\'enard$^{4}$,
                    \and
           Hossam Aly$^{8}$
          }
    \authorrunning{C. Toci et al.}
    
   \institute{European Southern Observatory (ESO), Karl-Schwarzschild-Strasse 2, 
              85748 Garching bei Munchen, Germany.\\
              \email{claudia.toci@eso.org}
         \and INAF, Osservatorio Astrofisico di Arcetri, 50125 Firenze, Italy    
         \and
             Dipartimento di Fisica, Università degli Studi di Milano,
             Via Celoria 16, 20133 Milano, Italy. 
         \and
             Univ. Grenoble Alpes, CNRS, IPAG, 38000 Grenoble, France.   
         \and
Department of Astronomy, University of California, Berkeley CA, 94720, USA.         \and
             Univ Lyon, Univ Lyon1, Ens de Lyon, CNRS, Centre de Recherche Astrophysique de Lyon UMR5574, F-69230 Saint-Genis-Laval, France
         \and
             Dipartimento di Matematica, Università degli Studi di Milano, Via Saldini 50, 20133, Milano, Italy
        \and
            Faculty of Aerospace Engineering, Delft University of Technology, Kluyverweg 1, 2629 HS Delft, The Netherlands
             }

   \date{Received November 2nd, 2023; accepted April 8th, 2024}

% \abstract{}{}{}{}{} 
% 5 {} token are mandatory
 
  \abstract
  % context heading (optional)
  % {} leave it empty if necessary  
   {GG Tau is one of the most studied multiple young stellar systems: GG~Tau~A is a hierarchical triple surrounded by a massive disc and its companion, GG Tau B, is also a binary. Despite numerous observational attempts, a comprehensive understanding of the geometry of the GG~Tau~A system is still elusive.} %Given the significant role of dynamical interactions in shaping the evolution of these systems, it is relevant to characterise the stellar orbits and the discs' properties. }
  % aims heading (mandatory)
   {To determine the best orbital configuration of the GG~Tau~A system and its circumtriple disc, we provide new astrometric measures of the system and we run a set of hydrodynamical simulations with two representative orbits to test how they impact a disc composed of dust and gas.}
  % methods heading (mandatory)
   {We test the dynamical evolution of the two scenarios on short and long timescales. We obtain synthetic flux emission from our simulations and we compare them with multi-wavelength observations of 1300 $\mu$m ALMA dust continuum emission and 1.67 $\mu$m SPHERE dust scattering to infer the most likely orbital arrangement.}
  % results heading (mandatory)
   {We extend the analysis of the binary orbital parameters using six new epochs from archival data, showing that the current measurements alone (and future observations coming in the next 5-10 years) are not capable of fully breaking the degeneracy between families of coplanar and misaligned orbits, but obtaining that a modest misalignment is probable. We found that the time-scale for the onset of the disc eccentricity growth, $\tau_{ecc}$, is a fundamental time-scale for the morphology of the system. Results from the numerical simulations show that the best match between the position of the stars, the cavity size, and the dust ring size of GG~Tau~A is obtained with the misaligned configuration ($\Delta\theta= 30^\circ$) on timescales shorter than $\tau_{ecc}$. The results exhibit an almost circular cavity and dust ring, favoring slightly misaligned ($\Delta\theta\sim 10 - 30^\circ$), low eccentricity ($e\sim 0.2-0.4)$ orbits. However, for both scenarios, the cavity size and its eccentricity quickly grow for timescales longer than $\tau_{ecc}$, and the models do not reproduce the observed morphology anymore. This implies that either the age of the system is shorter than $\tau_{ecc}$ or that the disc eccentricity growth is not triggered or dissipated in the system. This finding raises questions on the future evolution of the GG~Tau~A system and, more in general, on the time evolution of eccentric binaries and their circumbinary discs. 
   }
  % conclusions heading (optional), leave it empty if necessary 
   {}


   \maketitle
%
%-------------------------------------------------------------------

\section{Introduction}

Stars usually form in clustered environment \citep{Clarke+2000, Offner22}: the stellar formation process frequently leads to the formation of systems with two or more stars (\citealt{duchene2013stellar}).  This is in agreement with the measured high fraction of multiple stars in the early phase of star formation (between $40\%$ and $70\%$ in class 0 and I stars, \citealt{Chen2013,Tobin16}), and it is also found in numerical simulations of collapsing molecular clouds \citep{Bate, Bate2018}. Many of the known binaries host circumstellar or circumbinary discs (see e.g., \citealt{manara2019observational} for a sample in Taurus and the reviews from \citealt{Zagaria2023} and \citealt{Zurlo23}).
Given that the process of planet formation happens in protoplanetary discs, it is reasonable to imagine that a significant fraction of planets form in discs orbiting multiple stellar systems. Some of these planets have been recently detected (see e.g., \citealt{Doyle2011,Martin2019,Standing2023}), revealing a rich variety of exoplanetary architectures.

The final fate of the circumbinary(multiple) material around a binary(multiple) system is largely determined by the gravitational interaction with the central stars. For example, the tidal interaction between the stars leads to the truncation of the circumstellar discs of the systems, reducing their gas and dust mass, their sizes, and their lifetimes \citep{artymowiczlubow1994}.
This reflects on planet formation and evolution: different orbital parameters may generate a broad variety of planetary architectures, such as p-type circumbinary or circum-multiple planets (orbiting around the multiple stellar system barycenter) and s-type circumstellar planets (orbiting around one of the stellar components). Moreover, the planetary orbits could either be coplanar or misaligned with respect to the binary (or multiple) orbital plane. \citep{MoeKratter2021}.
 Theoretical and numerical results differ on the size of binary cavities on long timescales (see e.g., \citealp{thun2017circumbinary,hirsh2020cavity,ragusa2020evolution}), with numerical simulations predicting larger eccentric cavities. 
Moreover, several authors found a growth of the eccentricity of the binary and of the disc cavity in initially circular discs (e.g., \citealt{papaloizou2001orbital,Kley2006,Dangelo2006,dorazio2021}). Regardless of the initial eccentricity, Lindblad resonances are excited, driving a fast disc eccentricity growth \citep{ogilvie2003saturation}.  
Finally, for sufficiently high values of the mass ratio ($q \geq 0.5$), a crescent shape over-density orbiting at the edge of the cavity is a frequent outcome (e.g., \citealt{miranda2017viscous,ragusa2017,Poblete19}). 

The modelling of observed accretion discs in multiple stellar systems is a crucial test-bed for testing and improving our understanding of the disc evolution and planet formation processes in such a scenario.  For example, low  ($e<0.1$) to negligible disc cavity eccentricities are measured in some of the observed systems (e.g., CH Cha in \citealt{Kurtovic2022}, HD 98800 B in \citealt{Kennedy19}), while theory predicts eccentric cavities. In this context, the GG Tauri multiple system is a prime example to study. 
GG Tau is composed of two main components: GG~Tau~A, a hierarchical triple\footnote{i.e. a triple system consisting of an inner binary and an outer star, where the centre of mass can be closer to the inner binary or the outer star; for a review of the evolution of hierarchical stellar systems see e.g., \citealt{Triplereview}}, and its companion GG Tau B, itself a binary.
The source has been extensively studied in the literature and astrometric data has been collected for almost 20 years (e.g., \citealt{Luhman99, Beust2005}), but there is still great uncertainty regarding the orbital parameters of the stars, such as eccentricity, semi-major axis and eventual inclination between the disc and the multiple stars \citep{kohler2011orbit} (for a broader discussion, see Sec. \ref{sec:intro_GG}). 
These parameters are pivotal to study the stellar system-disc interaction, and therefore to explain the observed features. 

In the last few years, the dynamical evolution of the gas distribution in the GG~Tau~A system has been simulated by several authors, considering the system as a binary, with models differing in the inclination of the binary orbit with respect to the disc \citep{Nelson16,cazzoletti2017,aly2018secular,keppler2020gap}, but a final consensus on the best configuration has not been reached. No models including dust and gas dynamics have been post-processed to compare to disc observations.

In this work, we: i) provide new relative astrometric measurements for GG~Tau~A, incorporating additional points from archival data and ii) perform 3D SPH gas and dust hydrodynamic simulations of GG~Tau~A, looking for the conditions that better reproduce the multi-wavelength observations. 
In the next section, we present the GG~Tau~A system; 
in Section \ref{sec:astrometry} we describe the new GG~Tau~A astrometric data, and we infer the updated orbital properties of the stars. In Section \ref{sec:model} we describe the details of the set of models. We analyse the results for the coplanar and the misaligned cases in Sections~\ref{sec:copl_res} and \ref{sec:mis_res} respectively. The comparison with observations is given in Sec. \ref{sec:comparison}. A discussion is provided in Section~\ref{sec:discussion}. Finally, we draw our conclusions in Section~\ref{sec:concl}.

%--------------------------------------------------------------------
\section{GG~Tau~Aa/b}\label{sec:intro_GG}

GG Tau is a hierarchical quintuple system located in the Taurus-Auriga star-forming region at a distance of 145 pc \citep{Galli19}, with an estimated age between 1 and 4 Myr \citep{white1999test,kraus2009coevality}. 
The two main components, GG~Tau~A and GG Tau B, with a projected separation of about $\sim$ 10'' ($\sim$ 1500 AU; \citealt{leinert1991lunar, leinert1993systematic}), are also a triple (GG~Tau~Aa, Ab1, Ab2) and a binary (GG Tau Ba, Bb) systems themselves. In this work, we focus our attention on the northern and more massive system, the hierarchical triple system GG~Tau~A. The total dynamical mass of the system has been estimated in \citealt{phuong2020b}, and corresponds to $1.41 \pm 0.08$ M$_\odot$ after scaling to the adopted distance of this work.
The mass ratio between the primary and secondary components has been estimated in \citet{keppler2020gap} as $\sim 0.77$. 
The system is composed of the primary star, GG~Tau~Aa, and of a secondary object, GG~Tau~Ab, resolved with interferometric observations as a binary (GG~Tau~Ab1/Ab2) with total mass $M_{\rm ab}\sim0.8 M_\odot$ and a projected separation of about $26\pm 1$ mas ($\sim$ 3.6 au) (Duchêne et al., submitted). Given the small separation relative to the distance to the circumtriple disc, which is the subject of our study, we will consider GG~Tau~Ab1 and GG~Tau~Ab2 as a single component GG~Tau~Ab.

The system is surrounded by a massive disc. To date, we have access to a plethora of observations coming from dust thermal emission, scattered light emission, molecular line emission such as CO gas emission, and dust polarisation \citep{dutrey1994images,Silber2000,Krist2002,Phuon2021,Tang23}. 
The total disc mass and the disc inclination are $\sim$ 0.12 M$_{\odot}$ and  $i = 37 ^{\circ}$ with respect to the line of sight respectively \citep{guilloteau1999}. 
The gaseous disc extends out to more than $\sim$ 850 au and reveals a centrally cleared cavity, with prominent spirals. The origin of such spirals is still unclear; however, recent results invoke planet-disc interaction to explain their presence \citep{phuong2020b}. Scattered-light observations in the optical, near- and thermal infrared regime locate the inner edge of the circumtriple disc at about 190-200 au \citep{duchene2004multiwavelength}.
In addition, the dust distribution shows a ring shape: in fact, the population of large dust grains observed at (sub-) millimetric wavelengths is confined in a narrow ring surrounding a dust depleted cavity, peaking at 230-240 au \citep{andrews2014}. 
The ring appears smooth and homogeneous, with a localised azimuthal brightness variation below 20$\%$ \citep{phuong2020}. The circumstellar disc around the primary star, GG~Tau~Aa, has been detected at millimetre wavelengths, while the discs around the two other stars -- classified as a Classical T Tauri star in \citealt{White2001} -- are too compact to be detectable due to tidal truncation effects and radial drift \citep{phuong2020}.
The presence of a circumtriple dusty disc is indicative of dust particles being trapped within a pressure maximum at the edge of the cavity \citep{Pinilla2012}. SPHERE images in H band \citep{keppler2020gap} show a highly structured disc with an unresolved inner region, probably due to the presence of material around the three stars (GG~Tau~Aa, Ab1, Ab2); the latter could be also responsible for the shadows cast on the outer disc. In the cavity, filamentary-like structures are also detected, generally interpreted as accretion streams.

\section{Archival observations and orbit analysis}\label{sec:astrometry}

Since the orbit published by \cite{kohler2011orbit}, GG~Tau~A has been observed repeatedly with adaptive optics systems, providing a substantial increase in the orbital coverage of the  GG~Tau~Aa - (Ab1,Ab2) orbit. Searching through observatory archives, we identified six new epochs of 2\,$\mu$m imaging observations with Keck/NIRC2 and VLT/NaCo that extend the coverage by a full decade (see Table\,\ref{table:astrom}). We retrieved the NIRC2 fully calibrated data products from the Keck Observatory Archive\footnote{https://koa.ipac.caltech.edu/UserGuide/about.html}. From the ESO archive\footnote{http://archive.eso.org/eso/eso\_archive\_main.html}, we retrieved the raw dataframes and associated calibration files (dark and flat-field frames). The raw NaCo frames were dark-subtracted and flat-fielded. Both NIRC2 and NaCo frames were further background-subtracted, either using the median of a full sequence (in cases where the acquisition sequence employed dithering) or computing a frame-by-frame median background value. Finally, the individual images were registered and median-combined to produce final images for each epoch. The resulting images are diffraction-limited, with a FWHM of about 0\farcs045 and 0\farcs055 with NIRC2 and NaCo, respectively.

From these images, we proceed to evaluate the relative astrometry of the system through least squares fitting. Specifically, we minimised the difference between GG~Tau~Ab and a shifted and scaled down copy of GG~Tau~Aa. We performed this at the individual frame level and computed the standard deviation of the mean over all frames to evaluate astrometric uncertainties, which we quadratically combined with standard calibration uncertainty (i.e., on plate scale and orientation) to evaluate the random uncertainties. The results of this analysis are presented in Table\,\ref{table:astrom}, in which we also report the 2012 $K_s$ estimate from \cite{DiFolco2014} using the NaCo aperture masking mode for completeness. This methodology neglects the fact that GG~Tau~Ab is itself a close binary (GG~Tau~Ab1/Ab2) and effectively makes the generally incorrect assumption that the photocenter of that component is co-located with its centre of mass. The latter is necessary to fit the Aa--Ab orbit, but only the former is directly available in all epochs. The amplitude of this systematic error can be estimated as follows. The flux ratio of the system is  $\approx 0.2$ \citep[][Duch\^ene et al., submitted]{DiFolco2014}. Considering a very conservative range of  0.1--1 for the mass ratio of the Ab pair, one concludes that the error will be no larger than 30\% of the  instantaneous binary separation. Given that the orbital semi-major axis of the Ab1-Ab2 has been tightly constrained to 26$\pm$1\,mas (Duch\^ene et al., submitted), the systematic astrometric uncertainty is therefore at most,  and likely significantly smaller than, 10\,mas. 

\begin{table*}[htbp]
\centering
\begin{tabular}{c c c c c c c} 
 \hline\hline
 Obs. Date & Instrument & Filter & $N_{img} \times N_{coadd} \times t_{int}$\,(s) & Sep. (\arcsec) & PA (\degr) & Prog. ID \& PI / Ref.\\  
 \hline
 2010-12-09 & Keck/NIRC2 & Br$\gamma$ & 1\,$\times$\,1\,$\times$\,0.2 & 0.2522$\pm$0.0007 & 332.05$\pm$0.11 & U146N2L, A. Ghez \\ 
 2011-12-16 & Keck/NIRC2 & Br$\gamma$ & 9\,$\times$\,30\,$\times$\,0.4 & 0.2536$\pm$0.0007 & 330.88$\pm$0.20 & U054N2L, J.-L. Margot \\ 
 2012-10-29 & Keck/NIRC2 & $K'$ & 9\,$\times$\,30\,$\times$\,0.181 & 0.2558$\pm$0.0005 & 329.55$\pm$0.25 & H205N2L, J.Lu \\ 
 2012-12-06$^\dagger$ & VLT/NaCo & $K_s$ & (...) & 0.2558$\pm$0.0035 & 329.3$\pm$0.8 & \citet{DiFolco2014} \\
 2014-12-11 & Keck/NIRC2 & $K'$ & 9\,$\times$\,30\,$\times$\,0.181 & 0.2579$\pm$0.0005 & 327.48$\pm$0.11 & H233N2L, J.Lu \\ 
 2017-12-19 & VLT/NaCo & $K_s$ & 101\,$\times$\,1\,$\times$\,1.0 & 0.2557$\pm$0.0004 & 323.37$\pm$0.10 & 0100.C-0055, R. K\"ohler \\ 
 2019-10-13 & Keck/NIRC2 & Br$\gamma$ & 10\,$\times$\,20\,$\times$\,0.2 & 0.2553$\pm$0.0005 & 319.55$\pm$0.10\\ 
 \hline
\end{tabular} 
\vspace{0.2cm}
\caption{New astrometric measurements of the GG~Tau~A binary. The fourth column lists the number of independent individual images,  the number of individual frames coadded per image, and the individual integration time. No program ID or PI is listed in the Keck Observatory Archive for the last dataset. The $^\dagger$ symbol indicates the astrometric measurement from \citealt{DiFolco2014}, reported here for completeness.}
\label{table:astrom}
\end{table*}

The parallax to GG~Tau~A itself is severely affected by its underlying multiplicity and is unreliable \citep{Gaia16, Gaia23}. We therefore fixed the parallax of the system to 6.91\,mas, which is the mean estimate for the L1551 cloud \citep{Galli19}. A 2.3\% uncertainty is associated to this estimate. Furthermore, we applied floors of 0\farcs002 and 0\fdg5 on the uncertainties on separation and position angle, respectively, as a means to take into account the systematic error discussed above. The residuals from the orbital fits confirm that these amplitudes are appropriate.

To perform the fit, we employ two complementary methods and compare their results. On the one hand, we employed the {\tt orbitize!} package \citep{Blunt20} in combination with the parallel-tempered package {\tt ptemcee} \citep{Vousden16}. The free parameters used in {\tt orbitize!} are the eccentricity ($e$), inclination ($i$), phase of periastron ($\tau_0$, measured relative to the earliest data point), position angle of the line of nodes ($\Omega$), argument of periastron ($\omega$), semi-major axis ($a$), and total system mass ($M_{sys}$). From the latter two quantities, the orbital period ($P$) can be derived using Kepler's Third Law. As is standard for orbital fits, we employ uniform ($e$, $\omega$, $\Omega$ and $\tau_0$), log-uniform ($a$) and sine-uniform ($i$) priors. For the total system mass, we use a Gaussian prior based on the known dynamical mass of the system \citep[$1.41\pm0.08\,M_\odot$ after scaling to the adopted distance,][]{phuong2020}. We performed each fit with 10 temperatures and 100 walkers per temperature, advanced the chains 80000 steps, cut the first half as an extended burn-in, and only kept every tenth walker position to remove correlations in the chains. Inspection of the chain evolution confirms that they have reached convergence. The resulting cornerplots are presented in Appendix\,\ref{sec:appendix} and 68-percentile for all parameters are listed in Table\,\ref{tab:orbits}.

In a second method, we use the {\tt orbfit\_lib} procedure \citep{schaefer2006} that uses the fitting method developed by \cite{hartkopf1989} and \cite{mason1999}. This method relies on a random sampling of $P$, $e$ and $T_0$ (the time of periastron passage) within user-provided bounds, combined with a least square fit to the remaining parameters via the Thiele-Innes elements, and only keeping solutions that are within a given $\Delta\chi^2$ range of the best-fitting solution. This approach, or minor variations around it, are routinely used in fitting binary orbits \citep[e.g.,][]{Horch2021, Lester2022, Tokovinin2022}. Most relevant to this study, this is the same method that \cite{kohler2011orbit} to produce the last published orbit of GG~Tau~A. In addition to using the input astrometry, we specified broad ranges for the parameter space (i.e., orbital period and eccentricity ranging from 50 to 2000\,yr and 0 to 0.99, respectively). We further required that acceptable solutions had a total system mass in the 1.32--1.50\,$M_\odot$ range (see above). 

For all (MCMC and least squares) solutions, we compute a misalignment angle relative to the circumstriple ring. While the geometry of the latter is well determined \citep[e.g.,][]{phuong2020}, there are small uncertainties on both the inclination and position angle of the major axis ($\pm$1\degr\ and 2\degr, respectively). To compute the misalignment angle, $\Delta\theta$, we therefore use Equation\,1 from \citealt{Czek2019} in which we draw the angles describing the disc geometry from the appropriate Gaussian distributions. Note that, by definition, $\Delta\theta \geq 0$ and we therefore expect a bias away from perfectly coplanar, although we stress that this is an inherent feature of a situation where neither plane is perfectly known.

\begin{table}
\caption{Orbital parameters of the GG~Tau~A orbit. The second and third columns indicate the 68-percentile from the posteriors and the best-fitting model from the MCMC run with {\tt orbitize!}, while the fourth column is the best fit from the least squares fit with {\tt orbfit\_lib}. The next-to-last line represents the misalignment angle between the orbital plane and the circumtriple ring.}
\label{tab:orbits}
\centering
\begin{tabular}{cccc}
\hline
 & \multicolumn{2}{c}{{\tt orbitize!}} & {\tt orbfit\_lib} \\
 & 68-percentile & Max Like. & Lowest $\chi^2$ \\
\hline
  $P$ (yr) & 199\,$^{+67}_{-28}$ & 276 & 1883 \\
  $T_0$ & 2097\,$^{+33}_{-92}$ & 2048 & 2029 \\
  $e$ & 0.22\,$^{+0.06}_{-0.05}$ & 0.31 & 0.79 \\
  $a$ (au) & 38\,$^{+8}_{-3}$ & 49 & 174 \\
  $a$ (mas) & 263\,$^{+57}_{-24}$ & 341 & 1203\\
  $i$ (\degr) & 36\,$^{+6}_{-5}$ & 47 & 54 \\
  $\Omega$ (\degr) & 100\,$^{+19}_{-40}$ & 107 & 130 \\
  $\omega$ (\degr) & 90\,$^{+54}_{-15}$ & 142 & 173 \\
  $\Delta\theta$ (\degr) & 12\,$^{+11}_{-7}$ & 17 & 29 \\
  $\chi^2_\mathrm{red}$ & -- & 2.07 & 2.01 \\
\hline
\end{tabular}
\end{table}

\begin{figure*}[htbp]
\centering
\includegraphics[width=0.48\textwidth]{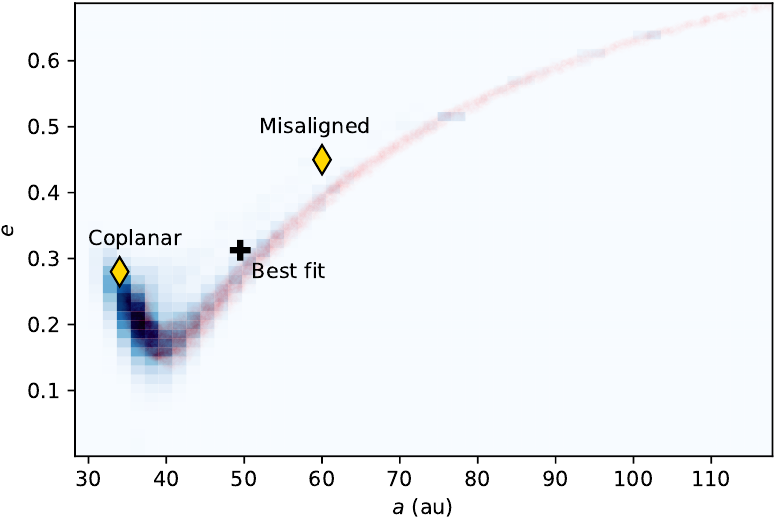}\hspace*{0.5cm}
\includegraphics[width=0.48\textwidth]{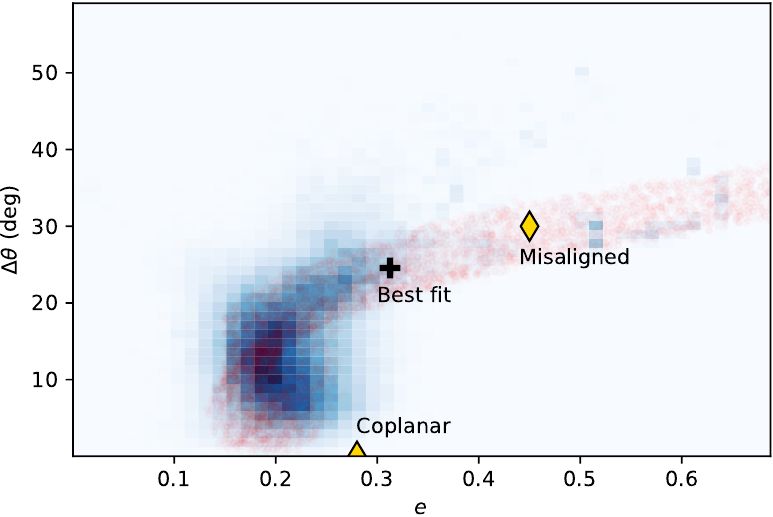}%
\caption{ Subset of the parameters describing the GG~Tau~A orbit: semi-major axis ($a$), eccentricity ($e$) and misalignment angle relative to the circumtriple ring ($\Delta\theta$). In both panels, the blue colormap represent the MCMC posteriors whereas red dots indicate solutions with $\chi^2 - \chi^2_\mathrm{min} \leq 1$ identified by the least squares fit. The red cross is the higest likelihood model from the MCMC chain and the yellow diamonds are the "coplanar" and "misaligned" solutions used in the hydrodynamical simulations of Section\,\ref{sec:model}.}\label{fig:orb_props}
\end{figure*}

Both the MCMC and least squares method yield overall good results. In particular, while the best fits have $\chi^2_\mathrm{red}\approx2.0$, 80\% of the deviations are incurred by the pre-2000 datapoints, which are the most likely to be affected by underestimated instrumental uncertainties, where the residuals for the post-2000 datapoints are consistent with well-estimated Gaussian errors. In line with \cite{kohler2011orbit}, our least squares fit finds that higher eccentricity orbits (which are also increasingly misaligned to the ring) are preferred from a $\chi^2$ perspective. However, the MCMC fit favors much less eccentric and misaligned solutions, albeit with the occasional excursion towards the "tail" of solutions from the least squares fit (see Figure\,\ref{fig:orb_props}). Indeed, the highest likelihood solution in the MCMC chain is outside the main posterior mode, a nod to the slightly better $\chi^2$ solutions. In short, high eccentricity solutions are formally more likely given the astrometric data, but they are much less plausible since they require the periastron passage happens within a very small fraction ($\lesssim0.1\%$) of the orbital phase. This analysis highlights the pitfalls of orbital fitting in situations where only short fractions of the orbit are available. As a general statement, the uniform sampling in $P$ and $T_0$ used in the least squares methods is inappropriate as it effectively oversamples the $\tau_0 \ll 1$ solutions, a problem that the MCMC setup avoids. Nonetheless, we should remain cautious in interpreting the results from the MCMC fit since the posteriors are complex, with low-probability tails that extend far from the primary modes (see Figure\,\ref{fig:orbitize_cornerplot}). 

Overall, it appears that our knowledge of the orbit of GG~Tau~A has not been dramatically improved despite the addition of nearly 10\,yr worth of orbital coverage. Nearly coplanar, low eccentricity ($e\approx0.2$) solutions are statistically more plausible, but more eccentric and misaligned orbits are more consistent with the astrometric datapoint. The most eccentric solutions found by the least squares fit can be ruled out based on the fact that their apoastron distance exceeds the semi-major axis of the circumtriple ring, which is implausible from a physical standpoint. Nonetheless, intermediate solutions, such as the "most plausible" orbit suggested by \cite{kohler2011orbit} and characterised by $e=0.44$ and $\Delta\theta=25\degr$, still appear fully consistent with the astrometric observations available to date. Since the peak of the posterior for $\Delta\theta\approx12\degr$ much exceeds the uncertainty on the orientation of the circumtriple ring, we conclude, albeit tentatively, that
a modest degree of misalignment is preferred by the currently available astrometric data. Further monitoring is required and it may be necessary to wait for more than a decade before significant progress on this front can be made, as the system will start approaching periastron. 

\begin{figure*}[htbp]
\centering
\includegraphics[width=0.48\textwidth]{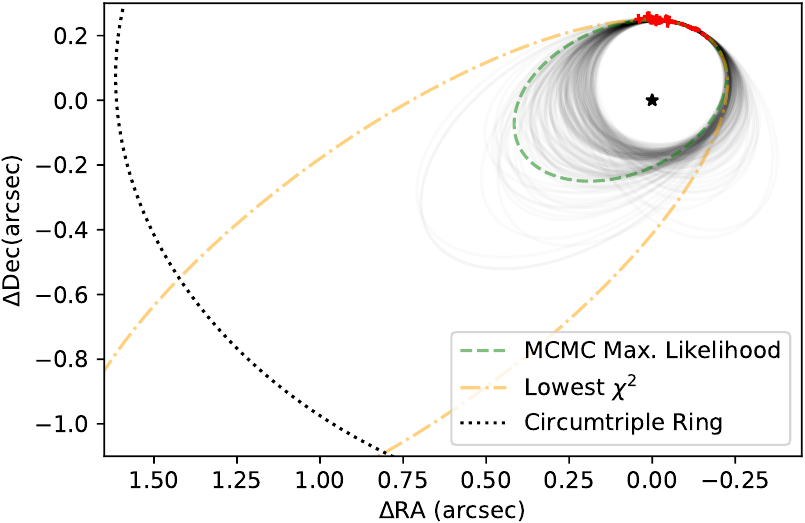}\hspace*{0.5cm}
\includegraphics[width=0.48\textwidth]{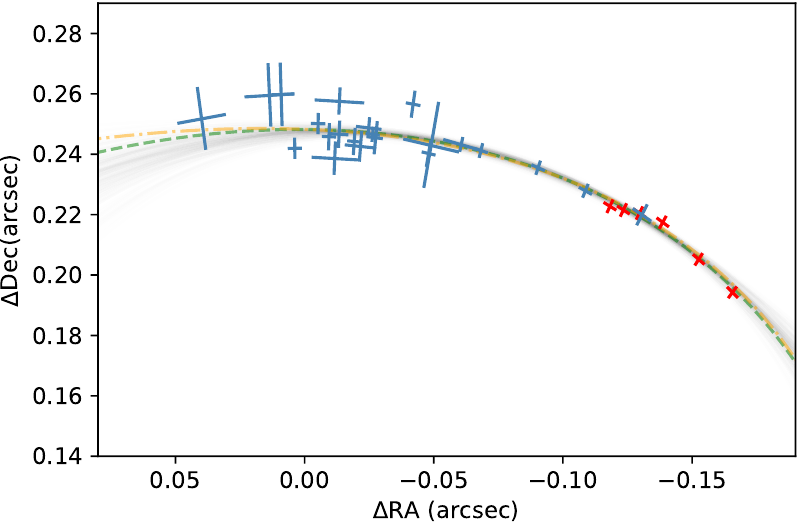}%
\caption{Left:  Random selection of 200 orbital solutions from the converged part of the MCMC chain. The green dashed and orange dot-dashed ellipses represent the highest likelihood model from the MCMC chain and the lowest $\chi^2$ orbit from the least squares fit, respectively.
The blue star marks the location of GG~Tau~Aa. whereas the dotted ellipse traces the continuum peak intensity along the ring \citep{andrews2014, Dutrey2014}. 
Right: Zoom on the region with astrometric coverage. Blue and red points indicate previously published and new astrometric measurements, respectively. The same orbits as in the left panel are rendered.}\label{fig:orbit}
\end{figure*}

\section{Numerical simulations}\label{sec:model}
As our new astrometric data points do not completely constrain the orbits of the system, we rely on the comparison between the morphology of the outcomes of numerical simulations and observations to identify the family of most likely configurations. We perform two sets of gas and dust 3D hydrodynamic simulations using the SPH (Smoothed Particle Hydrodynamics) code \texttt{PHANTOM} \citep{price2018phantom}, testing the effect of different initial conditions, and we then generate synthetic observations to compare with the observations directly.
While previous studies only considered the gas dynamics, our simulations include, for the first time in the context of this system, the gas-driven dynamics of the dust. This is a fundamental addition in order to test both the mm-emitting and the micron-scattering dust fluxes.
Indeed, among all the components observed, only mm-sized dust grains are optically thin and can therefore provide information on the density profile of large dust. 
We are also interested in probing the cavity size and eccentricity evolution, and if our models are capable of reproducing the observed streams of material detected with SPHERE images in the $H$ band, performing a multi-wavelength analysis. 

\subsection{Previous results}
We summarise here previous findings for bench-marking our gas dynamics results in the overlapping parameter space, ensuring accuracy in predicting the dust dynamics and the other timescales.
The tested configuration for the family of coplanar orbits has a semi-major axis of the binary of $a \sim 34$ au and an eccentricity of $e= 0.28$. 
The density (and thus pressure) maximum radial location at the cavity edge depends on the eccentricity and semi-major axis values \citep{artymowiczlubow1994}, which results in $\sim$ 100 au in the GG~Tau~A case if coplanar. 
This has been confirmed by numerical simulations of gas-only dynamics \citep{cazzoletti2017}, which also found that the pressure bump in the disc (where large dust is trapped, see \citealt{Pinilla2012}) is located at $\sim 150$ au, too compact with respect to observations that show a narrow ring at $\sim 220-240$ au.
The misaligned scenario has also been explored \citep{Nelson16,aly2018secular}. 
The plausible orbit tested
predicts a larger semi-major axis $a = 60$ au, an eccentricity of $e = 0.45$ \citep{kohler2011orbit}, while the inclination angle is poorly constrained. As described in Sec. \ref{sec:astrometry}, this orbit is still consistent with the updated astrometric analysis, and represents a good compromise between the best fit and all the allowed parameters' ranges (see Fig. \ref{fig:orb_props}).
For these values, the tidal truncation radius is also larger, $\sim 120-180$ au, and the maximum of the gas density should be located at $\sim 180- 220$ au.   Numerical results support this statement: \citealt{aly2018secular} evolved a set of gas simulations, finding that a misaligned binary with an initial inclination of $\sim 25^{\circ}$ produces a larger gas cavity, sufficient to explain the presence of a pressure maximum at the position of the dust observations. Recently, \citet{keppler2020gap} modelled again the system as coplanar, finding that on timescales larger than 1000 binary orbits, the binary excites in the disc a larger, eccentric cavity, hinting at the formation of a larger dust ring. However, at the state of the art, no models including dust and gas dynamics and their relative radiative transfer post-process have been explored. We aim to confirm these hypotheses by comparing theoretical results and multi-wavelength observations.

\subsection{Methods}
We tune our initial conditions to reproduce the main characteristics of the GG~Tau~A system, following the parameters choice of \citet{cazzoletti2017} and  \citet{aly2018secular} (see Table~\ref{table:1} and Fig.~\ref{fig:orbit} for a visualisation on our new astrometric analysis).
In particular, the two stars are modelled as sink particles \citep{bate1995modelling} with masses 0.78 M${_\odot}$ and 0.68 M${_\odot}$.
They exert gravitational forces on each other and on the gas and dust particles. In addition, they are subject to the back-reaction force caused by the gas, which ensures the overall conservation of the binary-disc angular momentum. 
Each of the two sink particles has an associated accretion radius, i.e. a radius within which we can consider gas and dust particles to be accreted onto the stars; in particular, we use $R_{\textrm{sink}}$ = 0.5 AU. We are aware that $R_{\textrm{sink}}$ is larger than the stars radii. Nevertheless, this choice is dictated by the fact that for our purposes we do not need to know what happens to the gas in the vicinity of the stars, so that we can consider gas particles at $R < R_{\textrm{sink}}$ as accreted. This approximation is not relevant for the dynamical evolution of the dust ring and the gas cavity, but may affect the morphology of the streams of material from the disc to the stars and the formation of inner discs \citep[e.g.][]{Ceppi+22}.
\begin{table*}
\centering
\begin{tabular}{c c c c c} 
 \hline\hline
 Orbital parameters & Coplanar 1 & Coplanar 2 & Misaligned 1 & Misaligned 2 \\  
 \hline
 Inclination $\Delta\theta$ [$^{\circ}$] & 0 & 0 & 30 & 30 \\ 

 Semi-major axis $a$ [au] & 34 & 34 & 60 & 60 \\

 Binary eccentricity $e$ & 0.28 & 0.28 & 0.45 & 0.45 \\
 Argument of periapsis $\omega[^{\circ}]$ & - & - & 270 & 0 \\
 
 Period T [yr] & 164 & 164 & 384 & 384 \\

 \hline
 \multicolumn{5}{c}{} \\ 
 \hline\hline
 Disc parameters & Coplanar 1 & Coplanar 2 & Misaligned 1 & Misaligned 2 \\  
 \hline

 R$_{\textrm{in}}$ [AU] & 68 & 34 & 120 & 120 \\

 R$_{\textrm{out}}$  [AU]& 800 & 800 & 800 & 800 \\  
 \hline
\end{tabular} 
\vspace{0.2cm}
\caption{List of parameters. \textit{Coplanar 1} and \textit{coplanar 2} differs for the value of R$_{\textrm{in}}$, \textit{misaligned 1} and \textit{misaligned 2} for the value of the $\omega[^{\circ}]$.}
\label{table:1}
\end{table*}

\subsection{Initial Conditions}

The system is surrounded by a disc of 1$\times 10^6$ SPH particles, corresponding to a resolved scale height <h>/H of $\sim 0.13-0.15$ for all the simulations. 
It extends between an inner radius R$_{\textrm{in}}$, which varies in the different configurations (see Table \ref{table:1} and Section \ref{sec:sim_set}), and an outer radius R$_{\textrm{out}}$ = 800 au and it is centred on the centre of mass of the binary. The surface density distribution is 
\begin{equation}\label{sigmaprof}
    \Sigma = \Sigma_0 R^{-p},
\end{equation}where $R$ is the radial coordinate in the disc, $p$ = 1 and $\Sigma_0$ is fixed to have a total disc mass M$_{\textrm{g, disc}}$ = 0.12 M$_\odot$  as in \citet{guilloteau1999}. For simplicity, we assume the standard initial dust-to-gas ratio value of $\epsilon=10^{-2}$, resulting in an initial dust mass M$_{\textrm{d, disc}} = 1.2 \times 10^{-3}$ M$_\odot$, assumed to be initially constant throughout the entire disc. This implies that the dust has initially the same vertical structure as the gas, but is allowed to vary during the simulation.  

Due to the highly expensive numerical cost of the simulation performed, we fix a single type of grains with size $a = 1$~mm and intrinsic density 3~g/cm$^3$.  For the selected set of parameters, the initial mid-plane Stokes number in all the simulations is smaller than unity. This justifies the use of the one fluid algorithm based on the terminal velocity approximation \citep{laibe2014dusty1, price2015fast}. Back-reaction from the dust on the gas is automatically included in this approach. We neglect the disc self-gravity, as our disc mass results in a gravitationally stable disc. 

We assume a locally isothermal equation of state for the gas, independent on the z for each radius, where the sound speed value $c_s$ varies as a power-law function of the radius with index $q=0.45$. We prescribe a radial temperature profile of the form
\begin{equation}
T(R)=20\rm{K}\left(\frac{R}{300 ~\rm{au}}\right)^{-0.9},
\end{equation}
as measured from $^{13}$CO observations (e.g., \citealt{Tang16}) and fixed in previous numerical studies \citep{cazzoletti2017,aly2018secular}. Assuming vertical hydrostatic equilibrium the aspect ratio is
\begin{equation}
    \frac{H}{R} = \frac{c_s}{v_k} = \Biggl(\frac{H}{R}\Biggr)_{\textrm{ref}}\Biggl(\frac{R}{R_{\textrm{ref}}}\Biggr)^{\frac{1}{2}- q},
\end{equation}
where  $(H/R)_{\rm ref}=0.12$ at $R_{\textrm{ref}}=100$ au to match our disc temperature. For misaligned simulations, the initial misalignment is  $\Delta\theta = 30^{\circ}$ as in \citet{aly2018secular}.

We set the value of $\alpha$ SPH artificial viscosity in order to have an effective \citet{ShakuraSunyaev} viscosity of $\alpha_{\textrm{SS}}$  = 0.005. To prevent particle interpenetration, we set the parameter $\beta$ = 2 as prescribed in \citet{price2012}.
In our initial conditions for the set of simulations, we do not include any circumstellar discs, as their numerical viscous dissipation times are extremely short and they do not affect the evolution of the circumtriple disc\footnote{The total angular momentum of the system is highly dominated by the angular momentum of the Aa-Ab pair, the effects produced by the binarity of Ab1-Ab2 appear only on secular timescales as perturbations to the Aa-Ab system on the disc.}, and their analysis is out of the scope of this paper.

\subsection{Simulations set}\label{sec:sim_set}

We run 4 SPH simulations with the \texttt{PHANTOM} code, two for the most probable coplanar configuration and two for the most probable misaligned one. 
Simulations for the same configuration differ in one unconstrained parameter only, to investigate the consequences of different initial conditions.
In particular, in the coplanar case, we explore the effect of two different initial conditions for the inner radius of the gas cavity: $R_{\textrm{in}}$ = [34; 68] AU. In the following, we refer to the simulation with the largest $R_{\textrm{in}}$ as \textit{coplanar 1} (which exactly replicates the setup from \citealt{cazzoletti2017}), and to the other as \textit{coplanar 2}. We note that in the former case, the inner boundary of the disc is equal to the binary semi-major axis $a$. 
According to the work of \citet{ragusa2020evolution}, this distinction in the initial condition can determine a different evolution for the gas for short timescales. Indeed, in their work, the authors find that a smaller initial inner radius leads to the formation of a pronounced, short lived over-density in the gas surface density. 

In the misaligned case a significant parameter is the argument of periapsis $\omega$, which measures the mutual inclination between the disc angular momentum  $\vec{j}$ and the binary orbital eccentricity  $\vec{e}$. The argument of periapsis corresponds to the angle from the body's ascending node to its periapsis. 
This parameter is very loosely constrained by observations. Thus, our choice is to perform a set of two simulations maximising the difference of values for $\omega = [0^{\circ}; 270^{\circ}]$ (corresponding to two orthogonal cases of $\vec{j}$ and $\vec{e}$).  
In the following, we will refer as \textit{misaligned 1} for the simulation with $\omega = 270^{\circ}$ and as \textit{misaligned 2} for  $\omega = 0^{\circ}$. In this case, we fix $R_{\textrm{in}}$  to 120 AU. The  \textit{misaligned 1} simulation replicates the best configuration found in \citet{aly2018secular}. 

In Table~\ref{table:1} we summarise the parameters used in our simulations. Each simulation is evolved until the system reaches a steady state. This is $\sim$ 3000 and $\sim$ 1500 binary orbits for the coplanar and misaligned cases respectively, which both correspond to a physical time of $\sim 5 \times 10^5$ yr.
\begin{figure*}%\vspace{-20pt}
\centering
\includegraphics[width=0.90\textwidth]{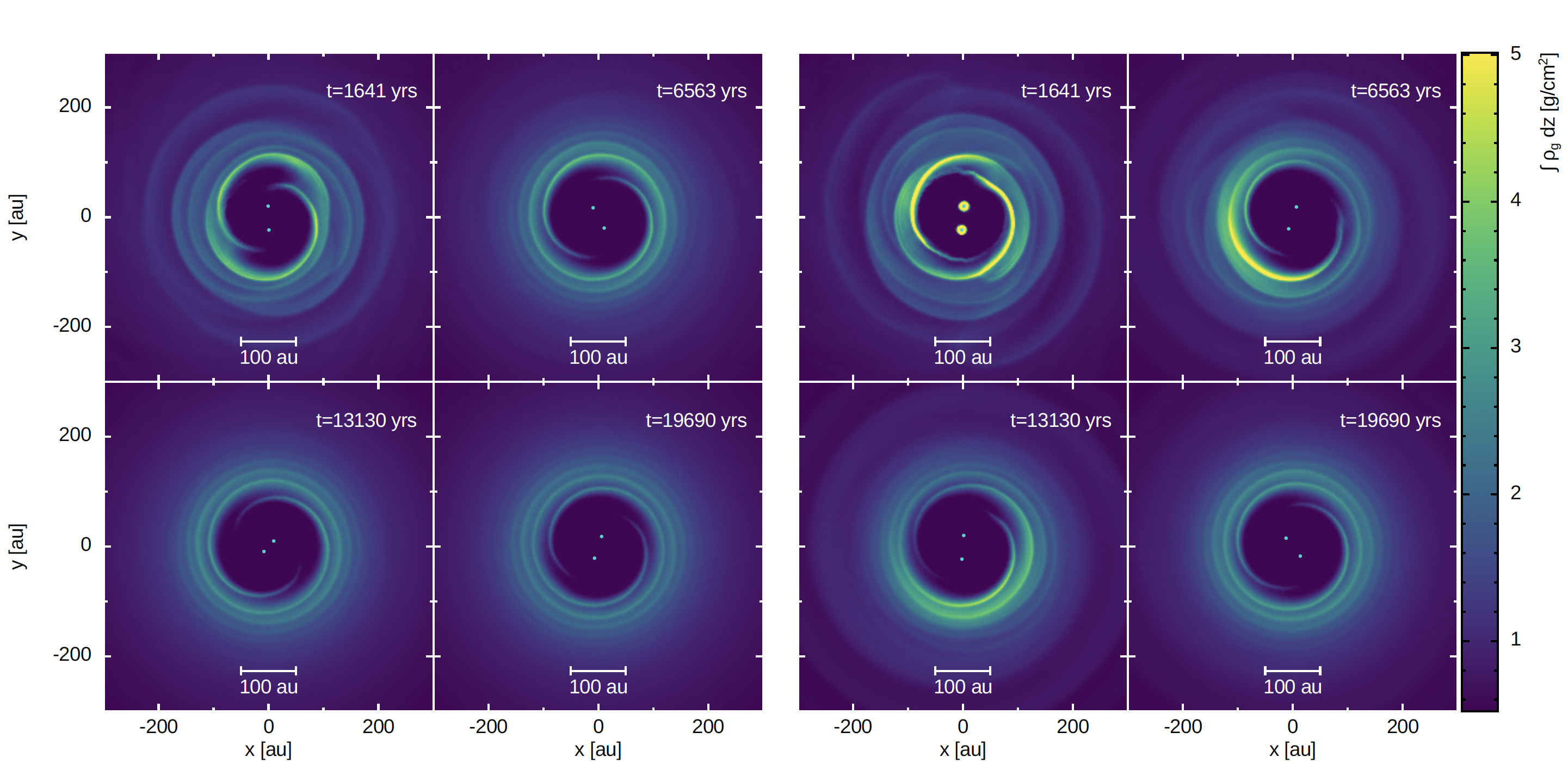}%\hspace{-5pt}
\caption{\textit{Coplanar} simulations, comparison between the initial behaviour shown with the same density and spatial scales. The cyan dots show the position of the binary. \emph{Left panels}: evolution of the gas surface density in the \textit{coplanar 1} simulation for face-on views at different orbital time steps, at 10 (top left), 40 (top right), 80 (bottom left), 120 (bottom right) T$_{\textrm{copl}}$.  The simulation is not forming an over-density in the gas. \emph{Right panels}: evolution of the gas surface density in the \textit{coplanar 2} simulation, at the same times. 
In both cases, the cavity remains circular. The initial snapshots of \textit{coplanar 2} also show how the initial accretion of material around the stars leads to the formation of circumstellar discs, quickly accreted due to low numerical resolution.}
\label{fig:coplanar_first}
\end{figure*}
\subsection{Radiative transfer and synthetic observations}\label{mcfostsetup}

We post-process the selected snapshots of all the simulations using \texttt{MCFOST} \citep{pinte2006monte}, a Monte Carlo radiative transfer code that maps physical quantities (such as dust density and temperature) directly from the SPH particles using a Voronoi tessellation.
The dust distribution in each cell is obtained with a linear interpolation in logarithmic scale of the grain size between 1~$\mu$m and 1~mm. Grains larger than 1~mm are assumed to follow the distribution of the 1~mm grains, while the ones smaller than 1~$\mu$m are assumed to follow the gas. We assume a power law grain size distribution $dn/da \propto a^{-3.5}$, with $a$ spanning in a range from $a_{\textrm{min}}$ = 0.03 $\mu$m and $a_{\textrm{max}}$ = 3.5 mm. As far as the optical properties are concerned, we use the effective medium theory (EMT) to derive the optical indices of the grains and the Mie theory (spherical and homogeneous dust grains) to compute their absorption and scattering properties. We assume a chemical composition of 60\% silicates, 15\% amorphous carbon, and 25\% porosity (in volume ratios), as in the DIANA standard dust composition \citep{woitke2016consistent}.  
The total gas and dust mass are directly taken from our SPH simulations. 

We consider the emission of the two stars as a black-body with an effective temperature T$_{\textrm{eff}}$ = 4078 K and 3979 K and a radius $R_\star$ = 1.55 $R_\odot$ and $1.5~R_\odot$ respectively, evaluated from the stellar masses according to the isochrones from \cite{Siess}. In all the simulations we fixed the distance of the source to $d$ = 145 pc, the inclination of the disc to $\Delta\theta = 37^{\circ}$ and we chose a position angle of PA = 97$^{\circ}$.  

Firstly, we compute the temperature of our models by using $1.3 \times 10^8$ photon packets in the Monte Carlo simulation assuming local thermal equilibrium (LTE). Then, we produce 1300 $\mu$m continuum images (corresponding to Band 6 ALMA observations) and 1.67 $\mu$m images of the selected snapshots, using the assumption that gas and dust have the same temperature. Finally, in order to generate a synthetic image that can be compared to real observations, we convolve the synthetic images with a Gaussian beam of the same size as the beam of the observations, 0.31'' $\times$ 0.25'', PA=27$^\circ$ for the 1300 $\mu$m emission (as in \citealt{phuong2020b}) and  0.040'' for the 1.67 $\mu$m emission (as the PSF value in \citealt{keppler2020gap}) respectively. 

\section{Results for the coplanar configuration}\label{sec:copl_res}

In the following Sections, we show the results of the two different sets of simulations, discussing the dynamical evolution of the systems on short and long timescales (indicatively, $< 1000$ T$_{\rm{orb}}$ and $\geq 1000$ T$_{\rm{orb}}$), and the synthetic images obtained with \texttt{MCFOST}. 
For the coplanar configuration, the binary period  T$_{\rm copl}$ will be used as time unit. For the chosen parameters T$_{\rm copl}$ corresponds to 1.64 $\times 10^2$ yr.

\subsection{Effect of the initial conditions of the inner edge of the disc}

\begin{figure}\vspace{-10pt}
    \includegraphics[width=.9\linewidth]{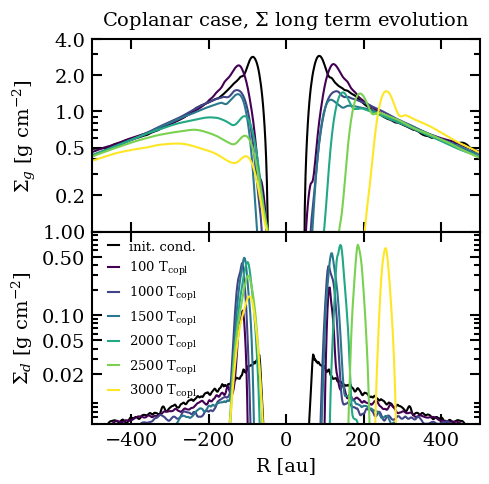}
\caption{\textit{Coplanar 1} simulation, radial cut along the major axis of the gas (top) and dust (bottom) surface density profile as a function of the radial distance from the star $R$ up to 3000 T$_{\textrm{copl}}$. After 1500 T$_{\textrm{copl}}$ the eccentricity of the gas cavity starts to grow. As a consequence, the dust ring moves further out.}\label{fig:sigmagasdustcomparisoncoplanar}
\end{figure}
The impact of the different initial conditions of the inner edge of the disc, reproducing the fact that the binary could or could not be initially embedded in the disc, is shown in Fig. \ref{fig:coplanar_first}, where we plot the evolution of the gas surface density for \textit{coplanar 1} (top panels) and \textit{coplanar 2} (bottom panels) in face-on views at $t=10, 40, 80, 120$ T$_{\rm copl}$. In the \textit{coplanar 2} case, an over-density forms at the edge of the cavity, precessing with Keplerian velocity, not present in the \textit{coplanar 1} case. At any rate, this difference is short lived. Indeed, the two different initial conditions considered eventually lead to the very same stationary configuration (see the bottom right panels of Fig. \ref{fig:coplanar_first}). 
Therefore, in the following we consider \textit{coplanar 1} as the representative configuration of the coplanar case in general.

\subsection{Time evolution}\label{sec:copl_time_short}

We let the \textit{coplanar 1} simulation evolve for $\sim 400$ orbits to relax the initial conditions, and then we monitor the dust and gas evolution for $\sim 1000$ more orbits, corresponding to $\sim 1.6\times 10^5$ yr. For a visualisation, Fig. \ref{fig:sigmagasdustcomparisoncoplanar} shows the evolution of the surface density profiles on a radial cut along the major axis of the gas (top) and dust (bottom). Our findings confirm the prediction that up to 1000 T$_{\textrm{copl}}$ the gas is truncated at $\sim$ 80 au, and we find that dust tends to collect at $\sim$ 100-120 au. Following \citealt{cazzoletti2017}, we select the snapshot at 1000 T$_{\textrm{copl}}$ as representative of the short timescale behaviour. The disc has a circular cavity, the dust and gas surface density do not show any evident asymmetry and the streams of material appear symmetric, as shown in Fig.~\ref{fig:cavitycoplanar_Cazz}.

\subsection{Long-term evolution}\label{sec:copl_time_long}
We evolved the \textit{coplanar 1} simulations up to 3000 orbital periods (corresponding to $\sim 5\times 10^5$ yr), confirming that long-time evolution processes may change the dust disc size. Indeed, from visual inspection of Fig.~\ref{fig:sigmagasdustcomparisoncoplanar} we note that the cavity grows in size and eccentricity, and we find that the maximum density moves to larger radii, with dust piling up in an eccentric ring with an apoastron distance of $\sim$ 200 - 250 au. 

In particular, after 2500 T$_{\textrm{copl}}$ (corresponding to $\sim 4.1\times 10^5$ yr), the cavity semi-major axis $a$ has grown in size up to $a \sim$ 140 au and it has become eccentric. We find that the dust ring forms at $\sim$ 200 au, which is closer to its predicted position in the model by \citet{andrews2014}. Note that a prominent over-density, with contrast ratio $\delta \sim 3-4$ \footnote{where $\delta$ is defined as the ratio between the density in the azimuthal feature and the density at the opposite side of the cavity}, forms in the gas and dust components, due to the progressive growth of eccentricity. This "eccentric feature" is caused by the clustering of orbits at their apocentres \citep{Athaiee2013,Teyssandier2016}.  

\begin{figure*}
\centering
\includegraphics[width=.8\textwidth]{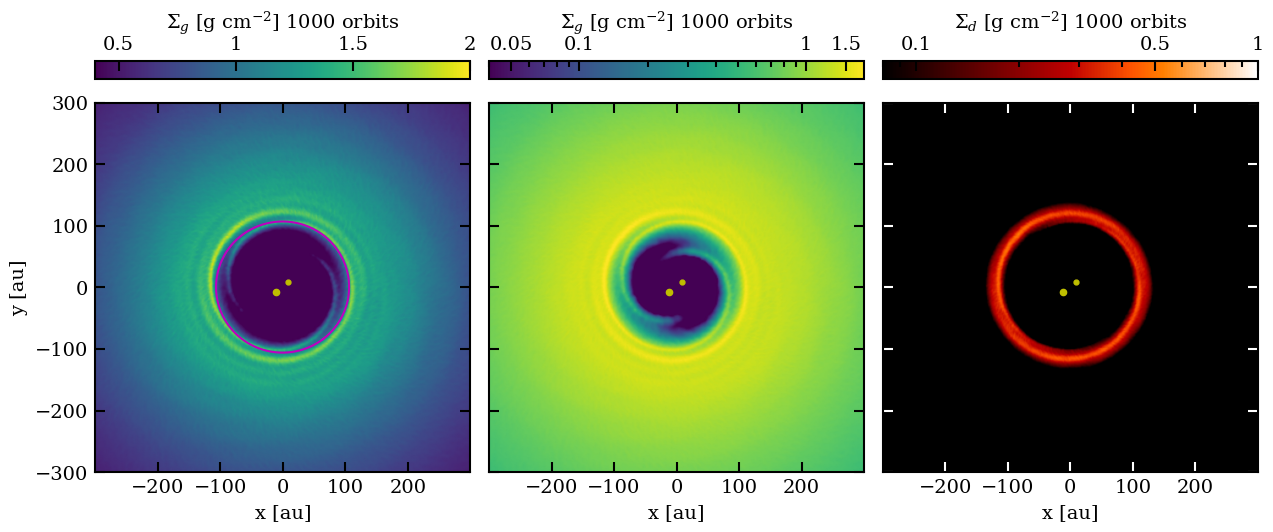}
\caption{\textit{Coplanar 1} simulation, Surface density after 1000 orbits, corresponding to $\sim 1.65\times 10^5$yr. The stars are shown as yellow dots. \emph{Left and Center:} Gas surface density after 1000 orbits (linear and log scale respectively). 
Symmetric accretion streams from the disc to the stars can be seen inside the gas cavity. The cavity size obtained with CaSh is over-plotted in red. \emph{Right:} Dust surface density. A circular dusty ring is formed at the location of the gas pressure maximum. No evident asymmetry is visible.}
\label{fig:cavitycoplanar_Cazz}
\end{figure*}

\begin{figure}%\vspace{-20pt}
    \includegraphics[width=0.9\linewidth]{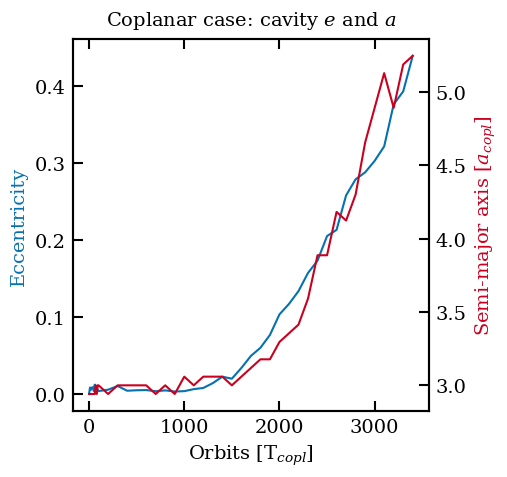}
\caption{\textit{Coplanar 1} simulation, evolution of the cavity semi-major axis $a$ expressed in units of the initial semi-major axis of the binary $a_{\rm{copl}}$, and evolution of the eccentricity $e_{\rm cav}$ as a function of time measured in T$_{\textrm{copl}}$. After $\sim$ 2000 orbits, $a$ and $e_{\rm cav}$ start to increase. 
}\label{fig:cavitysemimajoraxisevol} 
\end{figure}

\begin{figure*}
\centering
\includegraphics[width=.8\textwidth]{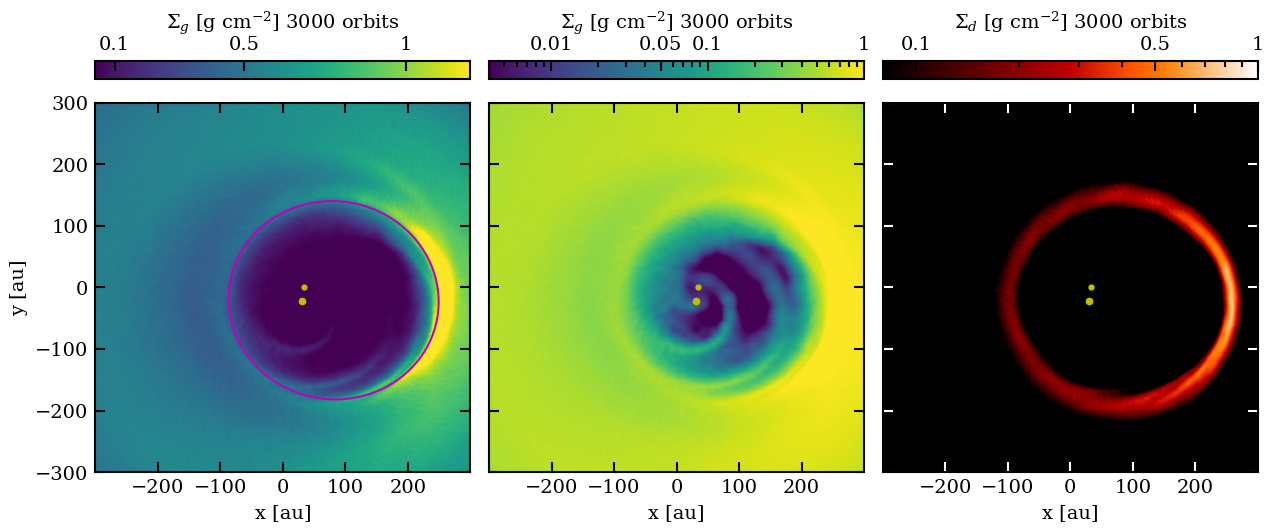}
\caption{ 
\textit{Coplanar 1} simulation, same quantities as Fig. \ref{fig:cavitycoplanar_Cazz} for the longer timescale (3000 orbits), corresponding to $\sim 4.9 \times 10^5$yr.  The long-time evolution leads to the growth of eccentricity in the disc. \emph{Left and Centre:} Gas surface density after 3000 orbits. \emph{Right:} Dust surface density. The dust is still trapped in the gas pressure maximum in an eccentric ring.  }
\label{fig:cavitycoplanar_final}
\end{figure*}

To quantify these results, we measure the semi-major axis $a$ and the eccentricity of the cavity $e_{\rm cav}$ in the gas component, using the numerical tool CaSh\footnote{CaSh is available at the following Github repository \url{https://github.com/si-mon-jinn/CaSh}.}. The results are shown in Fig. \ref{fig:cavitysemimajoraxisevol} (for a visualisation of our match with the cavity sizes, see Figs.~\ref{fig:cavitycoplanar_Cazz} and \ref{fig:cavitycoplanar_final}). For each simulation snapshot, CaSh computes the cavity eccentricity vector and semi-major axis as the average of the SPH particles contained in an interval around $R_{\textrm{mid}}$, and the semi-major axis of the inner cavity is defined as the radius $R_{\textrm{mid}}$, at which the surface density becomes half its maximum value \citep{artymowiczlubow1994}. In the following, we will refer to this quantity also as "cavity size". 

The cavity remains circular and of the same size (with a radius of $\sim 3$ $a_{\rm copl}$, corresponding to $\sim$ 100 au) for $\sim$ 1500 orbits, and then starts to grow in size and becomes increasingly eccentric. 
At the end of our simulations, after 3500 orbits, the cavity eccentricity has grown up to $\sim 0.4$, and the semi-major axis up to $\sim 5$ $a_{\rm copl}$, corresponding to $\sim 170$ au. The exact values of these quantities depend on the definition of the surface density threshold for the cavity size\footnote{Note that in this work, following \cite{artymowiczlubow1994}, the threshold for the surface density is 50$\%$ of the peak value, while in \cite{keppler2020gap} the threshold value is the 10$\%$ of the peak value. However, results are similar for the different thresholds.}. As the aim of the study is not to perform a study of the growth of $e_{\rm cav}$ in multiple discs, we decided not to evolve further the simulation in order to look for the convergence value, expected to be reached at later times \citep{ragusa2018eccentricity,hirsh2020cavity}. 
We adopt the 3000 T$_{\textrm{copl}}$ snapshot as representative of the "secularly evolved" scenario, which takes into account also long timescale processes, as it proves the best match to the eccentricity found by \cite{keppler2020gap}. The cavity size has grown to $a \sim 4.5$ $a_{\rm{copl}}$, with $e_{\rm cav} \sim 0.3$. The disc features a gas and dust over-density generated by the growth of eccentricity, and asymmetric streams, as shown in Fig.~\ref{fig:cavitycoplanar_final}. As a natural consequence of generating an eccentric cavity, the binary centre of mass is located in one of the foci. 

\begin{figure}
    \includegraphics[width=0.9\linewidth]{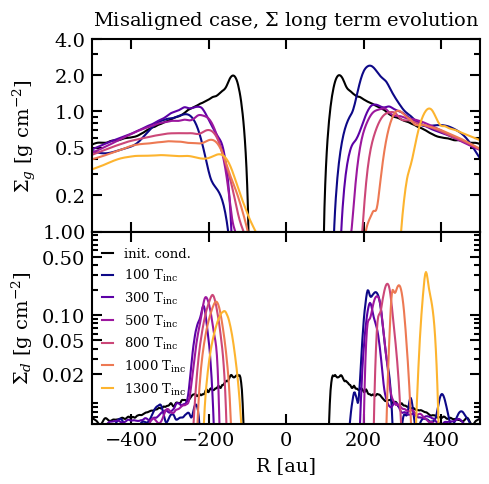}
 \caption{\textit{Misaligned 1} case, radial cut along the major axis of the surface density profile of the gas (top, $\Sigma_{\textrm{g}}$) and dust (bottom,  $\Sigma_{\textrm{d}}$) as a function of the distance of the star $R$ [au], for different time-steps. Initially, a low-eccentric ($e_{\rm cav} \lesssim 0.03$) cavity and dust ring form. Later on, the simulation shows a larger and more eccentric cavity and dust ring.} \label{fig:sigma_mis_comp}
\end{figure}

\section{Results for the misaligned configuration}\label{sec:mis_res}

In the following, we will use for the misaligned configuration the binary period T$_{\rm mis}$ as time units. For the chosen parameters T$_{\rm mis}$ corresponds to 4.3 $\times 10^2$ yrs.
\begin{figure*}
\centering
\includegraphics[width=0.8\textwidth]{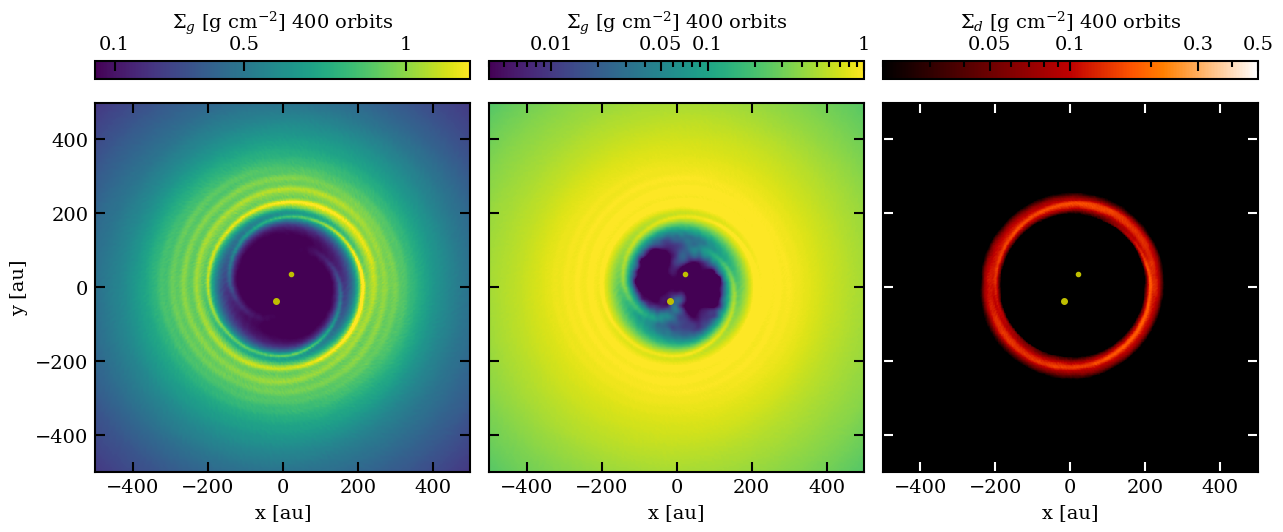}
\caption{\textit{Misaligned 1} simulation, surface density after 400 orbits, corresponding to $\sim 1.54 \times 10^5$ yr.  The stars are shown as yellow dots. \emph{Left and Centre:} Gas surface density after 400 orbits (linear and log scale respectively). \emph{Right:} Dust surface density. An almost circular dusty ring is formed at the location of the gas pressure maximum, larger with respect to the coplanar case size. }
\label{fig:cavity_mis}
\end{figure*}
\begin{figure}%\vspace{-20pt}
    \includegraphics[width=0.8\columnwidth]{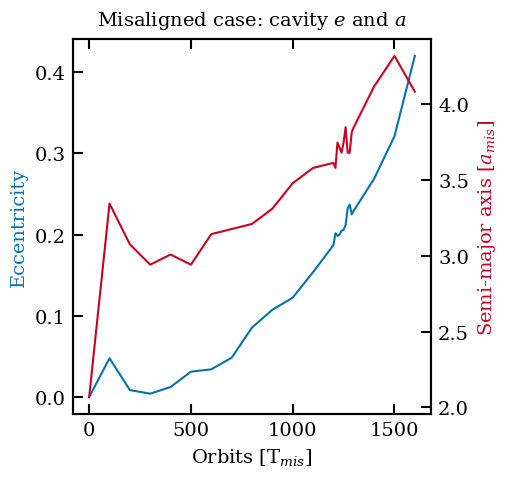} 
\caption{\textit{Misaligned 1} simulation, evolution of the cavity semi-major axis $a$ and the eccentricity $e_{\rm cav}$ as a function of the time, expressed in units of the period of the binary T$_{\textrm{mis}}$. The value of $e_{\rm cav}$ and $a$ are increasing after $\sim 600$ T$_{\textrm{orb}}$. Note that the initial cavity size jump is due to the stabilisation of the initial conditions. 
}\label{fig:Cash_mis}
\end{figure}

\subsection{Effects of the initial value of the binary argument of periapsis}

After checking that different initial conditions, once relaxed, do not impact the overall dynamics of the system, we concluded that the two cases evolve towards the same long-timescale behaviour. Indeed, the only difference between the two simulations is the position of the periastron of the system with respect to the chosen reference system, thus the position of the formation of the eccentric feature (see below).
Therefore, in the following, we consider the \textit{misaligned 1} configuration as representative of the misaligned case in general.  

\subsection{Time evolution}\label{sec:mis_time_short}

We let the  \textit{misaligned 1} simulation evolve for $\sim 200$ orbits to relax the initial conditions, and then we monitor the dust and gas evolution for $\sim 600$ more orbits (corresponding to $\sim 2.3\times10^5$ yr). Fig. \ref{fig:sigma_mis_comp} shows the time evolution of the gas and the dust surface density profile for the simulation. 
In agreement with theoretical and numerical results, the truncation radius (and thus the cavity size) is $\sim$ 120 au, with a pressure maximum located between 200 and 250 au. The latter corresponds to the position where dust tends to concentrate in our simulations. 
The gas cavity and the dust ring are also slightly eccentric, with $e_{\rm cav} < 0.03$, hardly noticeable from a visual inspection (see Fig. \ref{fig:Cash_mis} for the exact value). We select the 400 T$_{\textrm{mis}}$ snapshot to represent the "standard", short timescale behaviour of the misaligned configuration (as in \citealt{aly2018secular}), with an almost circular cavity of the size of 3 $a_{\rm{mis}}$, corresponding to $\sim 180$ au, and a symmetric dust ring and streams. The surface density profiles are shown in Fig. \ref{fig:cavity_mis}.

\subsection{Long term evolution}\label{sec:mis_time_long}

To test how long-time evolution processes impact also the misaligned case, we evolved for the first time the \textit{misaligned 1} simulation up to 1500 orbits (corresponding to $\sim 5.8\times 10^5$ yr). 
During the evolution of the simulation, the value of the inclination of the disc remains almost constant; at the same time, the disc starts to precess around the binary angular momentum. 

\begin{figure*}
\centering
\includegraphics[width=0.8\textwidth]{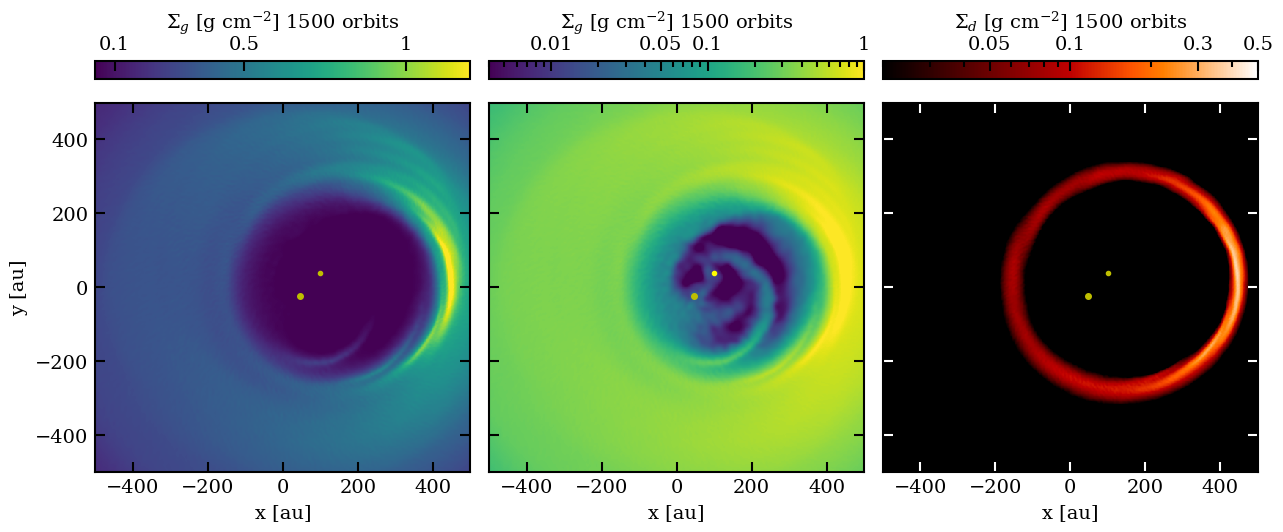}
\caption{\textit{Misaligned 1} simulation, surface density after 1500 orbits, corresponding to $\sim 5.7 \times 10^5$ yr.  The stars are shown as yellow dots. \emph{Left and Centre:} Gas surface density after 1500 orbits.  \emph{Right:} Dust surface density. The dust ring is now eccentric.  }
\label{fig:cavitymis_long}
\end{figure*}

From Fig. \ref{fig:sigma_mis_comp} it is clear that, also in the misaligned configuration the cavity size and eccentricity grow. In particular, after 1500 $\textrm{T}_{\textrm{mis}}$ (corresponding to $\sim 5.8 \times 10^5$ yr),  the cavity is larger, $\sim 240$ au, and it has become eccentric, with $e_{\rm cav} \sim 0.3$. As a consequence,  the dust ring becomes again wider and more eccentric, developing the eccentric feature with a contrast ratio $\delta=3-4$ at the apocentre of the orbits, as in the coplanar case. 
Fig. \ref{fig:Cash_mis} shows a quantitative analysis of the variation with time of the cavity parameters $e$ and $a$ for the \textit{misaligned 1} simulation.
We select the 1500 T$_{\textrm{copl}}$ snapshot to represent the "secularly evolved" scenario, visualised in Fig. \ref{fig:cavitymis_long}.

\section{Comparison with observations}\label{sec:comparison}
\subsection{Synthetic images}\label{sec:copl_synth_im}

In order to compare our simulations with ALMA and SPHERE observations, we post-process our representative snapshots to produce synthetic images at 1.67 and 1300 $\mu$m. 
The final gas mass for the two snapshots for the \textit{coplanar 1} case are 0.12 M$_\odot$ and 0.11 M$_\odot$ respectively, while the dust mass is $\sim 1.2 \times 10^{-3}$ M$_\odot$ for both snapshots. 
For the \textit{misaligned 1} case, the gas mass for the two snapshots is 0.11 M$_\odot$ and 0.095 M$_\odot$ respectively, while the dust mass is $\sim 1.18 \times 10^{-3}$ M$_\odot$. To perform a comparison between different simulations, we decided to fix the dust mass content in the radiative transfer post-process to $10^{-3}$ $ \rm{M}_\odot$, a value compatible with the gas mass estimate of the disc and gas to dust ratio of 100 \citep{guilloteau1999} and with the dust mass estimate from \citealt{andrews2014} ($\sim 10^{-3}-10^{-4}$ $\rm{M}_\odot$). 
\begin{figure*}
\includegraphics[width=\textwidth]{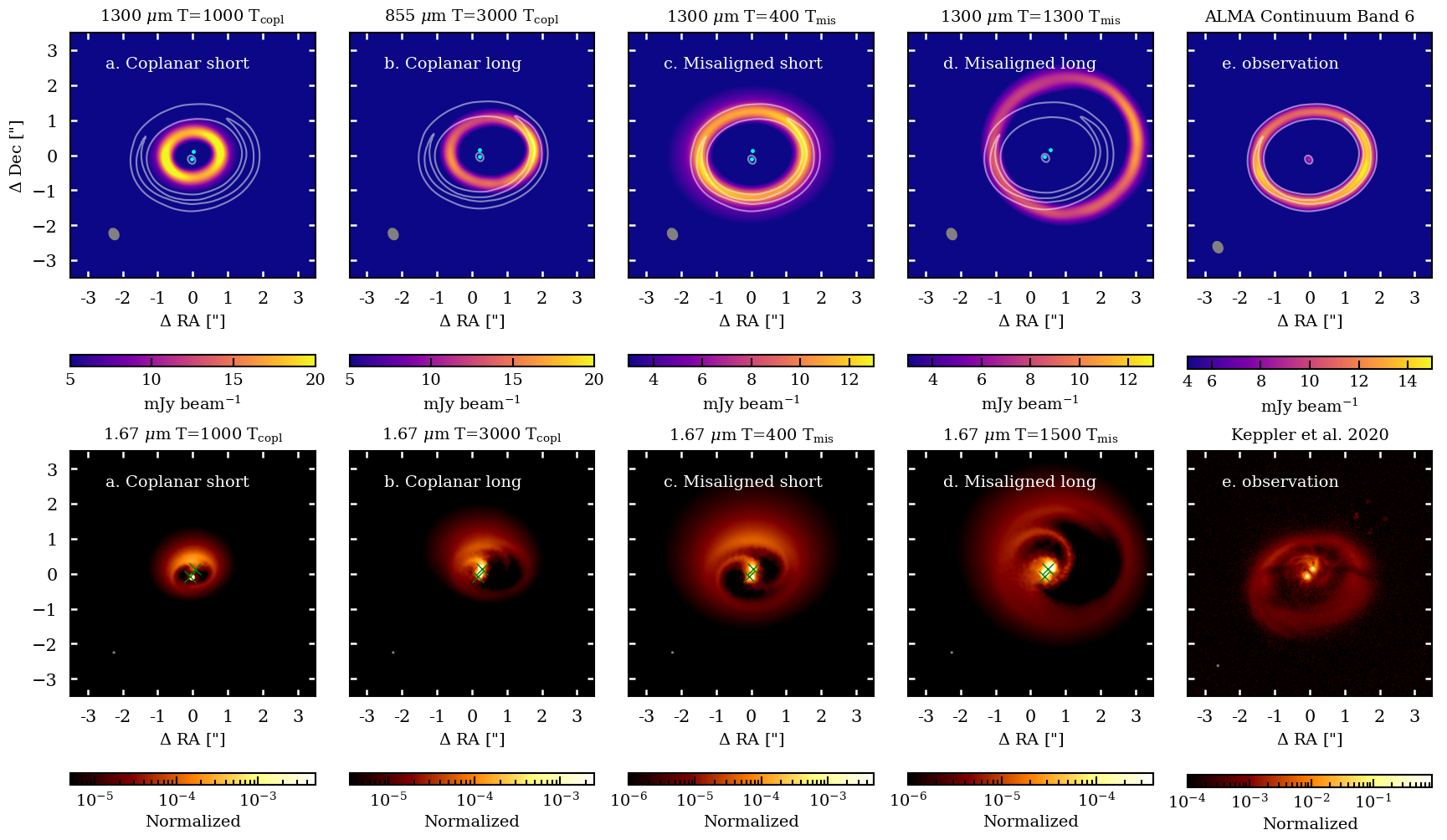}
\caption{Comparison between the synthetic images at 1300 $\mu$m (top) and 1.67 $\mu$m (bottom) from the four representative snapshots (two for the coplanar case, a. and b., and two for the misaligned case c. and d.) and the observed sizes of the dust ring of GG~Tau~A in Band 6 (see e.g., \citealt{phuong2020b}) and gas cavity obtained with SPHERE \citep{keppler2020gap}. 
The stars are shown as cyan dots. The 6 and 12 mJy contours of the continuum observation is over-plotted on the models in grey. In the continuum emission (top), the position of the circumstellar disc detected around GG~Tau~Aa is also shown, and it is used to centre the observed dust ring on the simulation. In the 1.67 $\mu$m images, the positions of the stars in \citealt{keppler2020gap} are shown as green crosses and are used to centre the observed scattered light image. } 
\label{fig:all_synth}
\end{figure*}
%%%
\subsection{Observed features}

The gas cavity size can be inferred by looking at the 1.67 $\mu$m image from \cite{keppler2020gap} (SPHERE/IRDIS observation, guaranteed-time observations GTO time, project 198.C-0209): assuming that the small dust, probed at this wavelength, is well coupled with the gas, the size of the cavity in the image is a good estimate of the cavity size of the system.

Likewise, information on the size of the large dust ring comes from the ALMA image of the continuum emission.  
In this work, we use as reference the archival ALMA Band 6 observations from project Project 2018.1.00532.S (P.I. Denis Alpizar, Otoniel), recently imaged and analysed in \cite{phuong2020b}, calibrated using the ALMA pipeline reduction scripts provided with the raw data from the ALMA archive.
Moreover, the model should also show prominent streams of material connecting the inner rim of the cavity to the stars, as observed in \cite{keppler2020gap}. Finally, the position of the stars with respect to the centre of the ring is another important signature of an eccentric cavity or ring. Indeed, while for a circular cavity the centre of mass of the binary is located in the centre of the circle, in the case of an elliptic cavity the mass centre of the binary is located in one of the foci. This difference can be in principle hidden due to projection effects, but can be tested in our synthetic images due to the fact that the inclination and the position angle of the GG~Tau~A system have been previously estimated (see Sec. \ref{sec:intro_GG}). In both observations, the position of the star is central with respect to the gas cavity and the dust ring, but the exact value of the eccentricity of the cavity and dust ring has never been measured. 
As a rough estimate, as the "centre offset" must be much smaller than the binary separation, $e \ll 0.2$. Hereafter, we will assume that the cavity is almost circular (i.e., with a low degree of eccentricity).  

In Fig.~\ref{fig:all_synth} we plot the synthetic images of the representative snapshots for the coplanar (panels a. and b.) and the misaligned (panels c. and d.) configurations, together with our reference images (top row, panel e., the ALMA Band 6 image and bottom row, panel e., the 1.67 $\mu$m image from \citet{keppler2020gap}). We also over-plot in grey the 6 and 12 mJy contours of the continuum observation on all the synthetic continuum images. To centre the observed and synthetic images in the same reference system, we used the position of the circumstellar disc of the primary star for the 1300 $\mu$m continuum observations, assuming that the primary star is at the centre of its circumstellar disc (top panels), and the estimated position of the stars from the accretion streams of the 1.67 $\mu$m image, shown as green crosses, matching the position of the primary star from the observations and from the model\footnote{the exact position of the secondary star may differ in each simulation, as it depends from the phase of the binary orbit and the relative inclination between the plane of the sky and the orbital planes.}. 

Visual inspection of the synthetic and observed 1.67 $\mu$m images of the system reveals that our models do no not fully reproduce the SPHERE morphology, with generally too bright and diffuse a northern (front) side. We believe that this indicates that, with our choice of dust opacity, the disc is either physically too thin or optically too thin in the simulations.
For this study, or main goal is to match the size of the inner cavity and not to produce a perfect match to all aspects of the ring. We therefore defer to a future work a more quantitative characterisation of the observed features.

\subsection{What is the most likely configuration for GG~Tau~A?}

Table \ref{table:2} shows a visual summary of our analysis. As discussed, the phenomenon determining the "short" and "long" timescale evolution in our simulations is the onset of the cavity eccentricity growth in the system, setting a threshold timescale $\tau_{ecc}$: the short- and long-time evolution can be then identified as T$_{orb} < \tau_{ecc}$ or T$_{orb} > \tau_{ecc}$ respectively, and the system shows different features at different timescales.
Focusing on the dust continuum emission (top panels of Fig. \ref{fig:all_synth}), we see that the main morphological features are not reproduced for the \textit{Coplanar} case at neither times (Fig. \ref{fig:all_synth} top panels a, b). 
Moreover, the model predicts a prominent emission asymmetry on the west side of the disc, not seen in any of the published results. 
Inspecting the \textit{Misaligned} configuration, we can see that the main features are well reproduced for the short timescale T$_{orb} < \tau_{ecc}$ snapshot (Fig. \ref{fig:all_synth} top panels c), but is not anymore a good match for longer timescales.
Considering the $\mu$m emission (bottom panels of Fig. \ref{fig:all_synth}), 
we find the same trend: the main features of the morphology of the cavity size are reproduced only in the \textit{Misaligned} case for T$_{orb} < \tau_{ecc}$.
We conclude that the misaligned configuration on short times is the best match of GG~Tau~A. This is no longer true for longer timescales, T$_{\rm mis} > \tau_{ecc}$.

\section{Discussion}\label{sec:discussion}
Our astrometric and hydrodynamical analysis leads to two main results: 

1)  The updated astrometric datapoints and orbital analysis presented here do not allow to conclude about the possible coplanarity of the circumtriple ring with the GG~Tau~Aa-Ab orbit, although a modest degree of misalignment appears likely.
%Our updated astrometric analysis indicates that the coplanar orbit is increasingly disfavored by the observed motion of the binary, suggesting that the binary-disc system has a misalignment. 
To consolidate this result, our multi-wavelength morphological analysis yields the same outcome.   
This implies that the GG~Tau~A multiple system is misaligned rather than coplanar. 
GG~Tau~A appears as the typical circum-multiple stellar disc, in which all observed discs have a moderate misalignment (neither coplanar nor polar configurations, see e.g., \citealt{Ceppi+23}, except for the case of very tight binaries as shown in \citealt{Czek2019}).
Determining the actual misalignment has an intrinsic value in itself, as several attempts to understand the degree of misalignment between the disc and binary populations and its physical implications have been made (e.g., \citealt{Czek2019}).

2) Our results show that the \textit{same} orbital configuration can be \textit{observed} with different morphological characteristics as a result of time evolution, leading to the fact that none of the models is a good match for the observed morphology of the system for all the timescales tested. The fact that disc eccentricity growth appears to be a common phenomenon prompts the following question: "why is the ring (almost) circular?".
In the following, we will try to answer this question.

\vspace{0.2cm}
\begin{table*}
\centering  
\vspace{0.2cm}
\begin{tabular}{c c c c c c}    
    \cline{2-5}
    \hline\hline
    \multicolumn{1}{c|}{} & Astrometry & Cavity size & Position of the binary  & Streams  \\
    \hline
    \textit{Coplanar} short & \xmark & \xmark & \cmark & \xmark  \\
    \hline
    \textit{Coplanar} long  & \xmark & \xmark & \xmark & \cmark \\
    \hline
    \textit{Misaligned} short & \cmark & \cmark & \cmark & \cmark \\
    \hline
    \textit{Misaligned} long & \cmark & \xmark & \xmark & \cmark  \\
    \hline
\end{tabular}
\vspace{0.2cm}
\caption{Table summarising our results from the comparison between GG~Tau~A models and observations. \textit{Coplanar} short and \textit{Coplanar} long refer to the results coming from the representative snapshots for the \textit{Coplanar} simulation at 1000 and 3000 T$_{\textrm{copl}}$ respectively, while \textit{Misaligned} short and \textit{Misaligned} long refer to the results coming from the representative snapshots for the \textit{Misaligned} simulation at 400 and 1500 T$_{\textrm{mis}}$. A green tick is shown where the feature is reproduced, while a red cross means that the feature is not reproduced.}         
\label{table:2} 
\end{table*}
\vspace{0.4cm}

\subsection{Eccentricity growth: the potential tension between numerical simulations and observations}
As already discussed in the previous Sections, eccentricity growth is in agreement with several numerical studies that found that a binary causes the disc eccentricity to grow (e.g., \citealt{papaloizou2001orbital,Kley2006,macfadyen2008,shi2012,miranda2017viscous,Munos2020,ragusa2020evolution,pierens2020,dittman2022,siwek2023,franchini2023}), and an overall consensus of binary evolution is found between all the most used (2D and 3D) numerical codes \citep{Codecomparison24}. Consistently with the previous literature, our two representative cases (i.e. coplanar and inclined) both develop cavity eccentricity growth. It is then reasonable to assume that other numerical models of this system will show eccentric cavities for \textit{most} choices of the system parameters if sufficiently evolved. How long each phase lasts is hard to know, with the only certainty that most works in Literature suggest that the final state will be eccentric. For this reason, we simply refer to "shorter" or "longer" timescales with respect to the onset of eccentricity growth, $\tau_{ecc}$.

In light of these considerations, it would be reasonable to expect eccentric circumbinary discs to be common in sufficiently evolved systems. However, this solid numerical result, where all circumbinary discs become eccentric, is somewhat in tension with the few observations available. Detecting the cavity shape of circumbinary discs and obtaining well-constrained inclination of these discs with respect to the binary orbital plane is a difficult task. Currently, we have well-constrained inclinations of only 10 circumbinary discs and 9 discs with inclinations measured with moderate uncertainties \citep{Czek2019,Ceppi+23}, but disc eccentricity has been measured only in a couple of sources (see below) and a handful of objects have resolved images of the disc, so their eccentricity can be at least visually estimated (6 in the sample of \citealt{Czek2019}, which explores in total 17 sources).

This small population of circumbinary discs with detected eccentricity (measured or at least visually estimated), consists of 9 systems with a range of ages, masses, and different binary orbital parameters: namely, 6 objects from the sample of \citealt{Czek2019}; IRAS~04158+2805 in \citealt{RagusaIRAS}; CS Cha in \citealt{Kurtovic2022}; GG~Tau~A in this paper). We find that 6 objects host almost circular cavities\footnote{V4046, AK Sco, DQ Tau, HD98800 B, GG~Tau~A, CS Cha, with 4 coplanar, 1 misaligned and one polar systems} and 3 objects host eccentric ($e_{\rm cav} > 0.1$) discs\footnote{GW Ori AB, HD142527, IRAS~04158+2805, all misaligned (and one triple system, GW Ori)}. This implies that the same issue raised in GG~Tau~A regarding the match between numerical expectations and observations is most likely to occur also when attempting the modeling of other systems characterised by circular cavities.

However, for completeness, we note that sources like IRS~48 \citep{Nienke2013,Calcino19,yang2023} or MWC~758 \citep{dong2018,kuo2022}, hosting eccentric cavities and asymmetric dust emissions, are often put in relationship with binaries, but lack the observational confirmation of the presence of a binary system. On top of that, the source HD142527 \citep{Casassus2013} exhibits an eccentric cavity and an asymmetric dust ring which can be explained by the presence of a binary \citep{pricecuello2018}. However, the detected companion appears to be too compact to explain the cavity size and shape \citep{Nowak24}.
Summarizing, this could possibly improve the statistics of the eccentric circumbinary discs, but would not solve the tension.

We finally note that an eccentric circumbinary disc may evolve in an eccentric planetary system if the disc eccentricity growth is not efficiently dissipated \citep{bitschKley10}. If \textit{all} the binaries would excite the disc eccentricity growth, we should then expect a higher occurrence of eccentric planets around binary stars. Results from ongoing surveys with ground (e. g., ESPRESSO) and space facilities (e.g., GAIA, TESS) will will help us to gauge the extent of these effects.

\subsection{Implications for GG~Tau~A and other circular circumbinary discs)}

In order to reconcile the fact that the cavity eccentricity growth appears to be an inevitable consequence in numerical works with the low eccentricity of the dust ring and the cavity observed in the GG~Tau~A system (but this applies also to other circumbinary discs observed to be circular), we identify two possible scenarios, and we discuss their implications:
\begin{itemize}
    \item 1. The growth of the cavity eccentricity is common for many cases but it simply does \textit{not} happen in GG~Tau~A.
    \item 2.  The growth of the cavity eccentricity will happen also in GG~Tau~A, but it has not occurred \textit{yet}.
\end{itemize}
In the first scenario, one needs to assume that the growth of the cavity size in eccentric binaries is a common and fast event, occurring for most of the systems on timescales shorter than the disc's lifetime. This implies that GG~Tau~A should be a special system. If this is the case, the multiple hierarchy between the components of the GG~Tau~A,B system may be responsible for damping eccentricity growth. The fact that GG~Tau~A is not a binary system but a hierarchical triple allows it to excite multiple oscillations modes  (see \citealt{Zanazzi17} or \citealt{Naoz16} for a review, and \citealt{Ronco21} for a specific effect that happens in multiple systems only). However, due to the compact scale of the secondary system, we believe that this effect is not the primary phenomenon responsible for the delaying or damping of the cavity eccentricity growth. Another intriguing explanation may be the presence of a circumtriple planet, located in the cavity, capable of slowing down the disc eccentricity growth \citep{Kurtovic2022}. Currently, no point source has been detected in the cavity, thus for the case of this system the presence of an inner planet is not yet supported with observations.  

The second scenario deals with the possibility that the GG~Tau~A disc will become eccentric in the future, but that the timescale for the onset of the cavity eccentricity growth has not been reached yet. This implies either that GG~Tau~A is a relatively young system if compared with the lifetime of systems hosting eccentric circumbinary discs ($\sim 1$ Myr) or that the timescale for the eccentricity growth is longer than GG~Tau~A's lifetime ($\sim 3$ Myr, see e.g., \citealt{white1999test}). However, suggesting that the system might be substantially younger would pose a considerable challenge without raising questions about the findings derived from evolutionary models. Therefore, this scenario is unlikely. 
A more promising way to slow down or damp the eccentricity growth could be the tidal interaction between the primary system GG~Tau~A, its circumtriple disc, and the secondary system GG Tau B.  
Indeed, previous studies found that the outer disc of GG~Tau~A is truncated \citep{dutrey1994images, Beust2005}, 
suggesting an efficient removal of material and, thus, angular momentum. If this is the case, other excitation modes could be removed, slowing down this kind of processes. 

By comparing our results for the \textit{Coplanar} case with \cite{keppler2020gap}, we see that $\tau_{ecc}$ in the two models, with similar initial conditions but different values of viscosity ($\alpha=$ 0.005 and 0.001 respectively), are different by a factor at least 2, with the shorter timescale associated with the higher viscosity. Also in the literature, the value of $\tau_{ecc}$ varies according to the different parameters of the simulations, but eccentricity growth is, as already mentioned, always happening.
This fact motivates us to speculate that the value of the viscosity plays a relevant role in determining $\tau_{ecc}$, possibly because viscosity is involved in regulating eccentricity or because it sets the timescale for the response of the system to the variation of its dynamics.
If so, its value is such that $\tau_{ecc}$ is so long (or that the growth is so slow) that the disc eccentricity growth will never happen.  A lower value of $\alpha$ with respect to the one tested (by at least one order of magnitude) may lead to an increase of $\tau_{ecc}$, but would also imply that we should not frequently observe very large and eccentric disc cavities.
Self-gravitating or cooling effects\footnote{as a higher cooling factor sets a decreasing of the thermal velocity of particles} --- here completely neglected --- may also be relevant in setting the angular momentum transport and the timescales of massive discs \citep{LodatoRice2005}.
However, as we currently lack a comprehensive understanding of how the actual physical mechanisms responsible for angular momentum transport influence the evolution of the disc's eccentricity, we might simply not fully understand eccentricity damping mechanisms. It is worth reminding that the $\alpha$ prescription is simply a parametrisation of the disc viscosity.

\section{Summary and conclusions}\label{sec:concl}

In this work we have analysed the GG~Tau~A system. 
We aim to find the best orbital configuration, namely if the binary-circumtriple disc system is coplanar or misaligned by combining multi-wavelengths information. We present new astrometric data points from Keck/NIRC2 and VLT/NaCo archival data and we perform an orbital fit, marginally improving the constraints on the orbital parameters. We then devise a suite of 4 numerical simulations with the 3D SPH hydrodynamical code \texttt{PHANTOM} \citep{price2018phantom}: two to study the coplanar configuration and two for the misaligned one, using the best configuration from the astrometric fit as orbital parameters. We tested the dynamical evolution of the two configurations on short and long timescales, and we included for the first time the large dust dynamics.

We showed that, whereas the orbital parameters cannot be determined from astrometry alone due to limited orbital coverage, a morphological analysis is required to obtain relevant information on the most probable orbits. Our main findings regarding GG~Tau~A are:
\begin{enumerate}
\item   We improve on the binary orbital parameters obtained in \cite{kohler2011orbit} by using a significant extended astrometric baseline and by comparing least squares and MCMC approaches to the orbital fit. We find that, despite the fact that the orbital coverage has increased by approximately 50$\%$ and all newly acquired data points exhibit high precision, the constraints on the orbital parameters have not undergone significant alterations when using the same least squares orbital fitting method. However, the MCMC approach shows that the lowest $\chi^2$ solutions are also much less plausible statistically speaking, highlighting the importance of priors in the fit and the need for an even longer baseline to make significant progress in constraining the orbit. All things considered, it appears likely, though not certain, than the ring is modestly misaligned with the orbit, and with low eccentricity.
\item For each configuration, we studied the effect of the initial condition, namely the initial cavity size for the coplanar configuration and the mutual inclination between the angular momentum of the binary and the cavity for the misaligned configuration, showing that those parameters do not influence the global evolution of the simulation. 
\item  We find that the coplanar configuration on short timescales has a cavity size and a dust ring size too small with respect to the observed ones. Instead, on longer timescales, this configuration leads to a larger and eccentric gas cavity with prominent accretion streams. However, it still fails to reproduce the basic observed features (cavity size, dust ring size). 
\item The misaligned configuration on shorter timescales is able to reproduce all the observed features of GG~Tau~A without requiring a highly eccentric gas cavity. We thus conclude that the most probable configuration for the GG~Tau~A system is a modestly misaligned configuration with binary-semi-major axis $a_{\rm mis}\sim 50 - 60$ au, eccentricity $e\sim 0.2-0.4$ and $\Delta\theta\sim 10^\circ - 30^\circ$. 
\item On longer timescales, once the cavity eccentricity growth modes are excited, the misaligned configuration also fails to reproduce the observations, as the gas cavity becomes too large and too eccentric. Such a fast growth of the binary eccentricity is at odds with the age of the system.
\end{enumerate}

The last result implies that: (i) the timescale for the onset of the eccentricity growth $\tau_{ecc}$ is a fundamental timescale for the system evolution; (ii) $\tau_{ecc}$ is longer than the age of the system; or (iii) there are physical mechanisms playing a
role in the disc eccentricity evolution that we are not considering. 
Due to the specific hierarchical configuration of the system, the tidal interaction between the GG~Tau~A circumtriple disc and the GG Tau B binary system (known to determine the outer disc truncation) may be responsible for slowing down the eccentricity growth.

In a broader context, our findings prompt us to consider if the onset of the eccentricity growth is a common process in the \textit{observed} binary systems, critically impacting their morphology and the architecture of the eventual hosted planetary systems. Eccentric cavities could be frequent and dust can potentially be trapped in the resulting over-density. This would lead to the formation of planets in eccentric circumtriple orbits. Moreover, as the binary in an eccentric cavity is located in one of the foci, this gives stringent constraints on the observational signatures of a binary in an eccentric disc. Ongoing and future surveys will be key in better constraining the population of eccentric discs and exoplanets in multiple stellar systems.

\begin{acknowledgements}
We thank the anonymous Referee for the useful suggestions. CT thanks Francesco Zagaria and Nicolas T. Kurtovic for the insightful scientific discussions and the valuable support.
This project has received funding from the European Union’s Horizon 2020 research and innovation programme under the Marie Sklodowska Curie grant agreement N. 823823 (DUSTBUSTERS RISE project).
This project has received funding from the European Research Council (ERC) under the European Union Horizon Europe programme (grant agreement No. 101042275, project Stellar-MADE).
This project has received funding from the European Research Council (ERC) under the European Union's Horizon Europe research and innovation program (grant agreement No. 101053020, project Dust2Planets).
E.R. acknowledges financial support from the European Union's Horizon Europe research and innovation programme under the Marie Skłodowska-Curie grant agreement No. 101102964 (ORBIT-D).
E.R. also acknowledges financial support from the European Research Council (ERC) under the European Union’s Horizon 2020 research and innovation programme (grant agreement No. 864965, PODCAST). H.A. acknowledges funding from the European Research Council (ERC) under the European Union’s Horizon 2020 research and innovation programme (grant agreement No 101054502).
This research has made use of the Keck Observatory Archive (KOA), which is operated by the W. M. Keck Observatory and the NASA Exoplanet Science Institute (NExScI), under contract with the National Aeronautics and Space Administration.

This paper makes use of the following ALMA data: ADS/JAO.ALMA 2018.1.00532.S ALMA is a partnership of ESO (representing its member states), NSF (USA) and NINS (Japan), together with NRC (Canada), MOST and ASIAA (Taiwan), and KASI (Republic of Korea), in cooperation with the Republic of Chile. The Joint ALMA Observatory is operated by ESO, AUI/NRAO and NAOJ. In addition, publications from NA authors must include the standard NRAO acknowledgement: The National Radio Astronomy Observatory is a facility of the National Science Foundation operated under cooperative agreement by Associated Universities, Inc. This paper uses the following Python libraries: \texttt{Numpy, Matplotlib, Astropy, Scipy, Glob}, and the following visualization tools: \texttt{pymcfost, plonk}.
\end{acknowledgements}

% WARNING
%-------------------------------------------------------------------
% Please note that we have included the references to the file aa.dem in
% order to compile it, but we ask you to:
%
% - use BibTeX with the regular commands:
%   \bibliographystyle{aa} % style aa.bst
%   \bibliography{Yourfile} % your references Yourfile.bib
%
% - join the .bib files when you upload your source files
%-------------------------------------------------------------------

%\bibliographystyle{aa}
\bibliographystyle{bibtex/aa}
\bibliography{GGtau}

\appendix

\section{Orbital fit: MCMC cornerplot}\label{sec:appendix}

\begin{figure*}
\includegraphics[width=\textwidth]{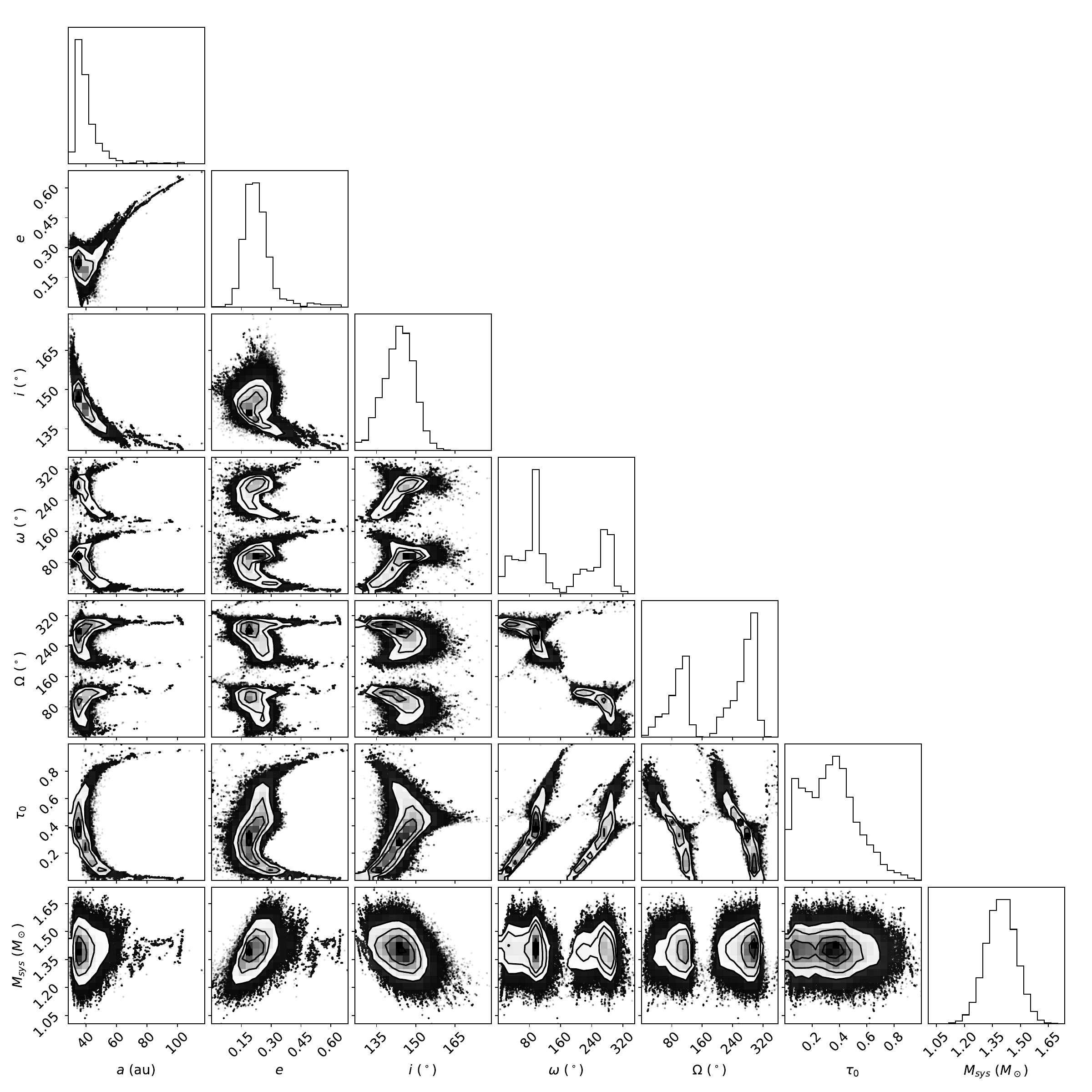}
\caption{ Cornerplot resulting from the {\tt orbitize!} fit to the GG~Tau~A orbit.} 
\label{fig:orbitize_cornerplot}
\end{figure*}

In Figure\,\ref{fig:orbitize_cornerplot}, we present the full cornerplot obtained using {\tt orbitize!} to fit the orbit of the GG~Tau~A system.

\end{document}